\documentclass{aa}

\usepackage{amssymb}
\usepackage{pifont}
\usepackage{graphicx}
\usepackage{xcolor}
\usepackage{amsmath}
\usepackage{pgfplots}
\usepackage{txfonts}
\usepackage{placeins}

\usepackage{natbib}                   
\bibpunct{(}{)}{;}{a}{}{,}

\usepackage[colorlinks=true,allcolors=blue]{hyperref}

\newcommand{\Hii}{H~{\sc ii}}

\begin{document} 

\title{Resolving dust and Ly$\alpha$ emission in a lensed galaxy at the epoch of reionization with JWST/CANUCS}
   \titlerunning{Dust and Ly$\alpha$ properties of LAE at z$\sim 6.6$}
   \author{V. Markov\inst{1}, M. Brada\v{c}\inst{1,2}, V. Estrada-Carpenter\inst{3,4}, G. Desprez\inst{5}, G. Rihtar\v{s}i\v{c}\inst{1}, J. Jude\v{z}\inst{1}, R. Tripodi\inst{6}, M. Sawicki\inst{7}, G. Noirot\inst{8}, N. Martis\inst{1}, C. Willott\inst{9}, R. Abraham\inst{10, 11}, Y. Asada\inst{10}, G. Brammer\inst{12, 13}, J. Matharu\inst{14}, A. Muzzin\inst{15}, G. T. E. Sarrouh\inst{15}, S. Withers\inst{15}, A. Ferrara\inst{16},  S. Fujimoto\inst{10, 11}, S. Gallerani\inst{16},   I. Goovaerts\inst{8}, A. Harshan\inst{17, 18}
     }
   \authorrunning{Markov et al.}
      \institute{Faculty of Mathematics and Physics, University of Ljubljana, Jadranska ulica 19, SI-1000 Ljubljana, Slovenia
         \and
         Department of Physics and Astronomy, University of California Davis, 1 Shields Avenue, Davis, CA 95616, USA
         \and
         School of Earth and Space Exploration, Arizona State University, Tempe, AZ 85287, USA
         \and
         Beus Center for Cosmic Foundations, Arizona State University, Tempe, AZ 85287, USA
         \and
         Kapteyn Astronomical Institute, University of Groningen, P.O. Box 800, 9700AV Groningen, The Netherlands
         \and
         INAF - Osservatorio Astronomico di Roma, Via Frascati 33, Monte Porzio Catone, 00078, Italy
         \and
         Department of Astronomy and Physics and Institute for Computational Astrophysics, Saint Mary's University, 923 Robie Street, Halifax, Nova Scotia B3H 3C3, Canada
         \and 
         Space Telescope Science Institute, 3700 San Martin Drive, Baltimore, Maryland 21218, USA
         \and
         National Research Council of Canada, Herzberg Astronomy \& Astrophysics Research Centre, 5071 West Saanich Road, Victoria, BC, V9E 2E7, Canada
         \and 
         Dunlap Institute for Astronomy and Astrophysics, 50 St. George Street, Toronto, Ontario, M5S 3H4, Canada
         \and
         David A. Dunlap Department of Astronomy and Astrophysics, University of Toronto, 50 St. George Street, Toronto, Ontario, M5S 3H4, Canada
         \and
         Cosmic Dawn Center (DAWN), Denmark
         \and
         Niels Bohr Institute, University of Copenhagen, Jagtvej 128, DK-2200 Copenhagen N, Denmark
         \and
         Max-Planck-Institut für Astronomie, Königstuhl 17, D-69117 Heidelberg, Germany
         \and
         Department of Physics and Astronomy, York University, 4700 Keele St. Toronto, Ontario, M3J 1P3, Canada
         \and
         Scuola Normale Superiore, Piazza dei Cavalieri 7, 56126 Pisa, Italy
         \and
         Kavli Institute for Cosmology, University of Cambridge, Madingley Road, Cambridge, CB3 0HA, UK
         \and
         Cavendish Laboratory - Astrophysics Group, University of Cambridge, 19 JJ Thomson Avenue, Cambridge, CB3 0HE, UK
             }
   \date{Received 15 12, 2025; accepted 25 02, 2026}

  \abstract
   {Lyman $\alpha$ (Ly$\alpha$) emission is highly sensitive to dust and neutral hydrogen and is expected to be suppressed in dusty or hydrogen-rich galaxies - especially during the epoch of reionization (EoR). Yet,  moderately dusty Ly$\alpha$-emitters (LAEs) are observed at this epoch, suggesting that complex interstellar medium (ISM) geometries and feedback-driven outflows may facilitate Ly$\alpha$ escape.
   }
   {We investigate the dust, gas, and stellar properties of the gravitationally lensed LAE HCM 6A at $z = 6.5676$ to characterize its multi-phase ISM structure and the physical conditions that regulate Ly$\alpha$ escape.}
   {We combine {\it JWST}/NIRISS slitless spectroscopy, {\it HST}+{\it JWST}/NIRCam imaging, and {\it JWST}/NIRSpec slit spectra from the CANUCS program. Using a customized \texttt{BAGPIPES} SED-fitting framework with a flexible attenuation law, we derive spatially resolved stellar, nebular, and dust properties on integrated ($\approx$ kpc), slit-level ($\approx0.1$ kpc), and pixel-level ($\approx25$ pc) measurements, thanks to strong lensing with magnification of $\mu \approx 8.3-9.1$. A high-quality Ly$\alpha$ map from \texttt{SLEUTH}, a tool for extracting spatially resolved emission-line maps from slitless spectroscopy, traces the spatial distribution of Ly$\alpha$ emission.} 
   {We measure an unlensed stellar mass of $\log M_\ast = 8.3$–$8.4$ and an intrinsic UV magnitude of $M_{\rm UV} = -19.8 \pm 0.1$.
   Slit-level measurements show that the older, more massive region (S1), is moderately dusty with consistent stellar ($A_V$, $\beta$) and nebular ($A_V^{\rm B}$, line-emission) indicators, implying a uniform ISM geometry, while the youngest region (S3) displays strong mismatches among dust tracers, revealing a complex, feedback-shaped multiphase ISM.
    Ly$\alpha$ emission arises primarily from S3. Pixel-level maps reveal a dust-cleared central clump (C3) where Ly$\alpha$ emerges, encircled by dustier outskirts, consistent with a very recent ($\lesssim 10$ Myr) starburst that created Ly$\alpha$ escape channels.
    Next, slit-level maps show Calzetti-like attenuation curves with a UV bump that increases with stellar age and decreases with $V$-band attenuation $A_V$, with a tentative detection of a UV bump in S1 at $\sim$25\% of the Milky Way amplitude. Pixel-level maps reveal that both the curve slope ($S$) and the UV bump ($B$) peak in the region between two clumps (C1 and C2), indicating dust-grain processing in a merger-driven starburst.
    }
   {Our observations provide a uniquely detailed, spatially resolved view of a moderately dusty LAE at the EoR, demonstrating how the interplay between multiphase ISM geometry and feedback governs Ly$\alpha$ escape. Constraining some key quantities requires fully resolved spectroscopy, which future JWST/NIRSpec IFU observations will provide.
   }

   \keywords{reionization, dust, extinction - galaxies: evolution  – galaxies: high-redshift - galaxies: ISM}

\maketitle

\section{Introduction} \label{intro}

The Lyman $\alpha$ (Ly$\alpha$) line at $1216 \AA$ is the strongest ultraviolet (UV) emission line in the spectra of star-forming galaxies. It arises from radiative transitions in hydrogen atoms, following either recombination or collisional excitation within \ion{H}{II} regions ionized by massive, young ($\lesssim 10$ Myr) O and B-type stars (e.g., \citealp{1967ApJ...147..868P, 2003A&A...397..527S}).  Because Ly$\alpha$ photons undergo resonant scattering with neutral hydrogen, the line profile and strength are susceptible to the presence, distribution, and kinematics of neutral hydrogen in both the interstellar medium (ISM) and the intergalactic medium (IGM; e.g., \citealp{1965ApJ...142.1633G}). In particular, absorption by neutral hydrogen in the partially neutral IGM gives rise to broad damping-wing absorption features that can suppress or asymmetrically distort the Ly$\alpha$ line profile, providing a key diagnostic of the IGM ionization state  (e.g., \citealp{2018ApJ...856....2M, 2024MNRAS.531L..34K, 2024MNRAS.532.1646K}). Consequently, the number of Ly$\alpha$-emitting galaxies (LAEs) declines toward the epoch of reionization (EoR), making them powerful probes of the ionization state of the IGM and the progress of cosmic reionization in the early Universe (\citealp{2004MNRAS.349.1137S,  2010Natur.468...49R, 2013ApJ...775L..29T, 2014ApJ...794....5T, 2018ApJ...856....2M, 2020MNRAS.495.3602W, 2023A&A...678A.174G, 2024MNRAS.531L..34K, 2024MNRAS.531.2998B, 2024A&A...688A.106N, 2025arXiv250918302P}; see \citealp{2014PASA...31...40D, 2020ARA&A..58..617O} for reviews). 
However, during the EoR, LAEs, often found in overdense regions, are expected to ionize their surroundings, forming the first ionized bubbles that, once expanded sufficiently, allow Ly$\alpha$ photons to redshift out of the line resonance and escape (\citealp{2018ApJ...856....2M, 2022MNRAS.515.5790L, 2024A&A...682A..40W, 2025ApJ...988...26W, 2025arXiv251018946N}). 

In addition, Ly$\alpha$ photons are subject to absorption and scattering by interstellar dust (e.g., \citealt{1999ApJ...518..138H}). While resonant scattering by neutral hydrogen primarily redistributes photons in wavelength and direction around Ly$\alpha$, dust absorption reduces the total photon number, thereby lowering both the Ly$\alpha$ escape fraction ($f_{\rm esc}^{\mathrm{Ly}\alpha}$) and the equivalent width (EW) of the line (\citealp{2008A&A...491...89V, 2009A&A...506L...1A, 2010ApJ...711..693K, 2011ApJ...730....8H, 2017ApJ...844..171Y, 2019MNRAS.486.2197B}).
Consequently, high-redshift ($z \gtrsim 4$) LAEs are generally characterized by low $V$-band dust attenuation (${A_V}_* \lesssim 0.5$), steep UV slopes ($\beta \lesssim -2$), young stellar ages ($\lesssim 5$ Myr), and low metallicities ($Z \lesssim 0.05 \ Z_{\odot}$; \citealp{2010ApJ...724.1524O, 2020MNRAS.493.5120M, 2024A&A...689A..44T, 2025ApJ...988...26W}).

However, both empirical studies and radiative transfer models demonstrate that even in moderately dusty systems ($A_V \lesssim 0.5 -1$), Ly$\alpha$ photons can escape efficiently through low-opacity sight lines within complex, multiphase ISM structures. In such configurations, Ly$\alpha$ photons scatter off the surfaces of clumps of neutral hydrogen gas and dust, and eventually escape through the ionized medium (\citealp{1991ApJ...370L..85N, 2006MNRAS.367..979H, 2008ApJ...678..655F, 2009ApJ...691..465F, 2009ApJ...704L..98S}). Additionally, stellar feedback and large-scale outflows play a dual role in facilitating Ly$\alpha$ escape: (1) they displace dust and neutral gas from the central regions, thereby reducing the optical depth, and (2) by accelerating neutral medium in an expanding shell, so that Ly$\alpha$ photons scattered off this outflowing gas emerge redshifted, reducing their subsequent resonant scattering and enabling escape (\citealp{2003ApJ...588...65S,2008A&A...488..491A}). Therefore, the interplay between ISM geometry and outflow-driven clearing facilitates Ly$\alpha$ escape despite the presence of dust (\citealp{2016ApJ...826...14G, 2021MNRAS.505.1382M}; see \citealt{2014PASA...31...40D} for a review).

LAEs exhibiting non-negligible dust attenuation during the EoR provide unique laboratories for exploring the interplay between Ly$\alpha$ emission and dust attenuation. Spatially resolved observations of such systems enable the study of local variations in the ISM on sub-galactic scales - including its multiphase structure, geometry, and feedback-driven outflows.

Building a comprehensive picture of this interplay and the physical processes governing it requires accurate constraints on the fundamental physical properties of LAEs, including the shape of the dust attenuation curve. Constraining the shape of the dust attenuation curve is essential both for gaining insight into the dust grain properties, content, and geometry, and for deriving reliable galaxy physical properties (e.g., stellar mass, star formation rate (SFR), stellar age) from SED fitting, which depend strongly on the assumed curve (e.g., \citealp{2013ApJ...775L..16K, 2015ApJ...806..259R, 2016ApJS..227....2S, 2023A&A...679A..12M}).

Spatially resolved studies of stellar, nebular, and dust properties uncover local physical conditions and processes within a galaxy that are otherwise washed out in integrated measurements. They reveal variations in dust and gas content, ionization state, stellar populations, star-forming clumps, feedback signatures, and merging subcomponents, which is particularly important for interpreting the formation and evolution. Recent observational JWST studies (e.g., \citealp{2024MNRAS.532..577E, 2025ApJ...991..188E, ormerod2025detection2175aauvbump}) alongside cosmological simulations combined with dust radiative transfer (RT;  \citealp{2026arXiv260207347N}) have demonstrated the power of such resolved mapping at the EoR.

In this paper, we present a case study of a well-known lensed, clumpy, luminous LAE at $z = 6.5676$ (\citealp{2002ApJ...568L..75H, 2005ApJ...635L...5C, 2007A&A...475..513B, 2013ApJ...771L..20K, 2020ApJ...896..156F}), that benefits from 19 photometric {\it HST} and {\it JWST}/NIRCam bands, three {\it JWST}/NIRSpec slits, and NIRISS imaging and spectroscopy from the Canadian NIRISS Unbiased Cluster Survey (CANUCS; \citealp{Willott_2022, 2026ApJS..282....3S}). We exploit spatially resolved maps of various dust diagnostics, attenuation curve properties, Ly$\alpha$ emission, and fundamental galaxy properties (stellar mass (${M_*}$), SFR (averaged over 10 Myr), mass-weighted stellar age (${\langle a \rangle}_*^{\rm{m}}$), ionization parameter ($\log{U}$), metallicity ($Z$), and the $V$-band attenuation ($A_V$)), comparing them with global trends and probing local variations in the environment (e.g., dust and gas content, stellar populations, and burstiness).

This paper is organized as follows. In Sect. \ref{Data}, we describe the CANUCS {\it JWST} observations and present our target galaxy. Sect. \ref{Method} details the SED-fitting procedure and the construction of Ly$\alpha$ emission maps from the NIRISS slitless spectroscopy. Our main results are presented and discussed in Sect. \ref{Results}. Finally, Sect. \ref{Summary} provides a summary and conclusions. Throughout the paper, we assume a $\Lambda$CDM cosmology  with $\Omega_{\rm m}=0.3$ and Hubble constant $H_0=70{\rm\ kms^{-1}\:\mbox{Mpc}^{-1}}$ for ease of comparison with previous work.

\section{Data} \label{Data}

To constrain the Ly$\alpha$ emission, dust attenuation curve, and fundamental galaxy properties of our source, we make use of the {\it JWST} NIRISS slitless spectroscopy, NIRSpec PRISM spectroscopy, and NIRCam imaging from the CANUCS Survey (\citealp{Willott_2022, 2026ApJS..282....3S}). The survey targets five massive lensing clusters, including Abell 370, to exploit gravitational lensing to probe faint, high-redshift galaxies. The survey combines coordinated {\it JWST} observations with NIRCam and NIRISS, and with a NIRSpec follow-up, in both cluster (CLU) fields and one flanking (NCF) field. The survey design and strategy are described in more detail in \cite{Willott_2022} and \cite{2026ApJS..282....3S}. 
 HCM~6A lies in the Abell 370 field at $\mathrm{RA}=39.9780111^\circ$, $\mathrm{Dec}=-1.558961^\circ$, in the outer region of the CANUCS/NIRCam pointing, in a region of significant gravitational lensing with a magnification factor of $\mu \simeq 8.3-9.1$.

\subsection{NIRISS slitless spectroscopy} \label{NIRISS_data}

The CANUCS observations (Program ID 1208, PI C. Willott) include NIRISS imaging and slitless spectroscopy of Abell 370. NIRISS imaging provides continuous coverage from $\sim 1-2 \ \mu m$ with 3.41 ks and 2.28 ks exposures in F115WN, F150WN, and F200WN, respectively ('N' is appended to each NIRISS filter to distinguish it from NIRCam filters), supplemented by deeper 3.86 ks F090WN imaging in three fields, including Abell 370, observed with the JWST in Technicolor program (ID 3362, PI A. Muzzin). Data processing procedure is fully described in \cite{2026ApJS..282....3S}. 

We use NIRISS Wide Field Slitless Spectroscopy (WFSS) observations obtained with the low-resolution ($R \sim 150$) GR150C and GR150R grisms crossed with all wide-band filters, including F090WN from Technicolor. The F090WN extends the wavelength coverage down to $\sim0.8\rm{\mu m}$, where Ly$\alpha$ falls at the redshift of our target galaxy ($z\approx 6.6$; \citealp{2002ApJ...568L..75H}).

The reduction of the NIRISS WFSS data largely follows the methodology described in \cite{2023MNRAS.525.1867N}, with the main difference that we use the CANUCS photometric catalog as input to \texttt{grizli} \citep{grizli23} to define source positions and segmentation maps, and to derive the spectral trace locations.
Basic detector-level processing follows the same approach adopted for the NIRCam imaging.
All subsequent steps — including astrometric alignment, background subtraction, flat-fielding, contamination modeling, and spectral extraction — are performed using \texttt{grizli} with the '221512.CONF' NIRISS configuration files (\citealp{matharu_2022_7628094}).
For the WFSS background subtraction, we adopt the commissioning background models; updated background templates that correct residual artifacts at the few-percent level \citep{2025jwst.rept.9057N} will be explored in future work.
A full description of the NIRISS imaging and WFSS observations, together with the complete data reduction procedure, will be presented in Noirot et al. (in prep.).

\subsection{NIRCam photometry} \label{NIRCAM_data}

Abell 370 NIRCam CLU field imaging spans 8 filters: F090W, F115W, F150W, F200W, F277W, F356W, F410M, and F444W, covering $0.9-4.4 \ \mu m$) with 6.4 ks exposures, using a 6-point dither pattern to fill detector gaps. 
The Abell 370 cluster was also observed with four additional NIRCam medium-band filters: F360M, F430M, F460M, and F480M from the JUMPS program ({\it JWST} Ultimate Medium-band Photometric Survey); ID: 5890; PI: Withers, Muzzin). 
The resulting dataset provides deep, multi-band coverage with well-characterized filter depths, enabling high-quality color imaging of the cluster field. {\it JWST} data were supplemented with existing {\it Hubble Space Telescope} ({\it HST}) Advanced Camera for Surveys (ACS) and the Wide Field Camera 3 (WFC3) bands in the CLU field (\citealp{Lotz_2017, Postman_2012, Steinhardt_2020}, (ID: 11507, PI: K. Noll).
The full list of available filters in CLU and NCF fields of Abell 370 is provided in Table 2 of \cite{2026ApJS..282....3S}.
A total of 19 photometric bands was used in our analysis, excluding the NIRISS filters since they probe similar wavelength ranges as the NIRCam bands.

Data reduction was performed using the CANUCS photometric pipeline, as described in detail by \cite{2026ApJS..282....3S}. This pipeline includes custom corrections for noise, persistence, cosmic rays, and artifacts. Images are astrometrically aligned to Gaia data release 3 (DR3), mosaicked with \texttt{grizli} (\citealp{grizli23}), and calibrated to a flux scale of 1 nJy. Bright cluster galaxies and intracluster light are modeled and subtracted following \cite{2018ApJS..235...14S} and \cite{martis24}. The end products are clean mosaics optimized for faint background galaxies in strongly lensed cluster fields.

\subsection{NIRSpec spectroscopy}

Follow-up spectroscopy was conducted with NIRSpec using the Micro-Shutter Assembly (MSA), as part of the CANUCS program (ID: 1208). Target selection was based on NIRCam and NIRISS imaging. To account for quadrant gaps, Abell 370 was observed with three MSA configurations in the CLU field, with each configuration observed for 2.9 ks using nodded 3-shutter slitlets. 

NIRSpec data processing follows the procedures described in \cite{2024MNRAS.530.2935D}, \cite{heintz25}, and \cite{2026ApJS..282....3S}. Stage 1 of the STScI {\it JWST} pipeline is applied, incorporating custom corrections for snowballs and 1/f noise, followed by Stage 2, which includes photometric calibration. Further reduction included \texttt{grizli} (\citealp{grizli23}) and \texttt{msaexp} (\citealp{msaexp}) packages, with wavelength calibration corrected for intra-shutter offsets, standard nodded background subtraction, and optimal 1D extraction accounting for PSF (point spread function) variation with wavelength. 

\begin{figure*}[h]
    \centering
    \includegraphics[width = 0.22\textwidth]{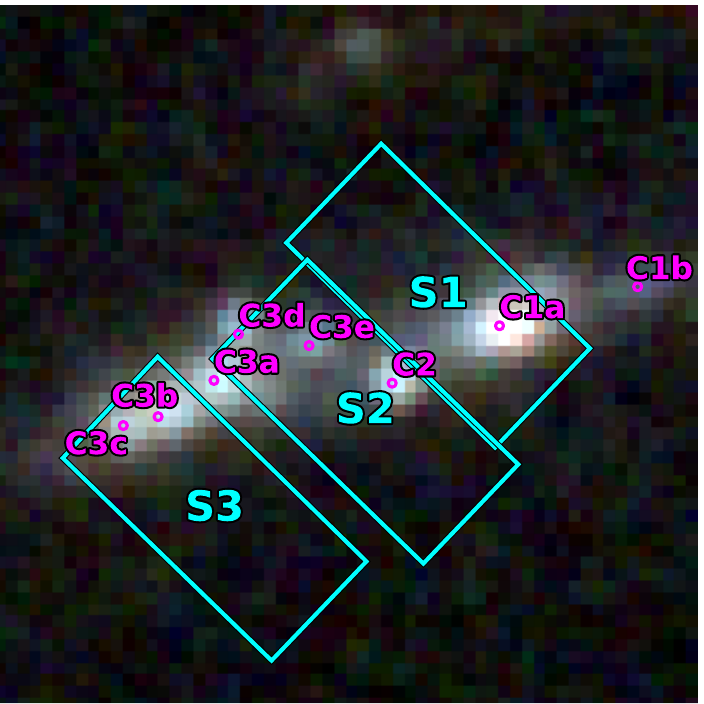}
    \includegraphics[width = 0.22\textwidth]{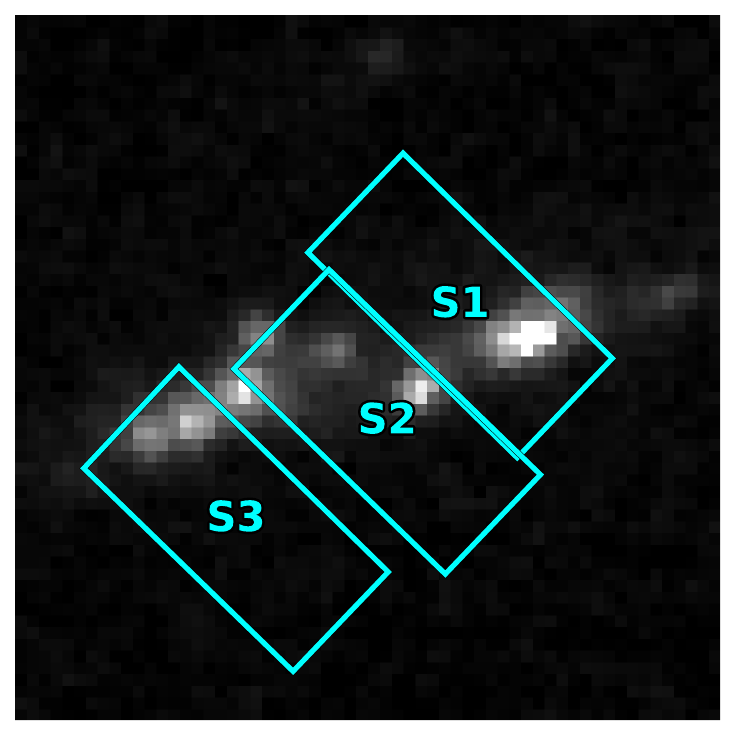} 
    \includegraphics[width = 0.49\textwidth]{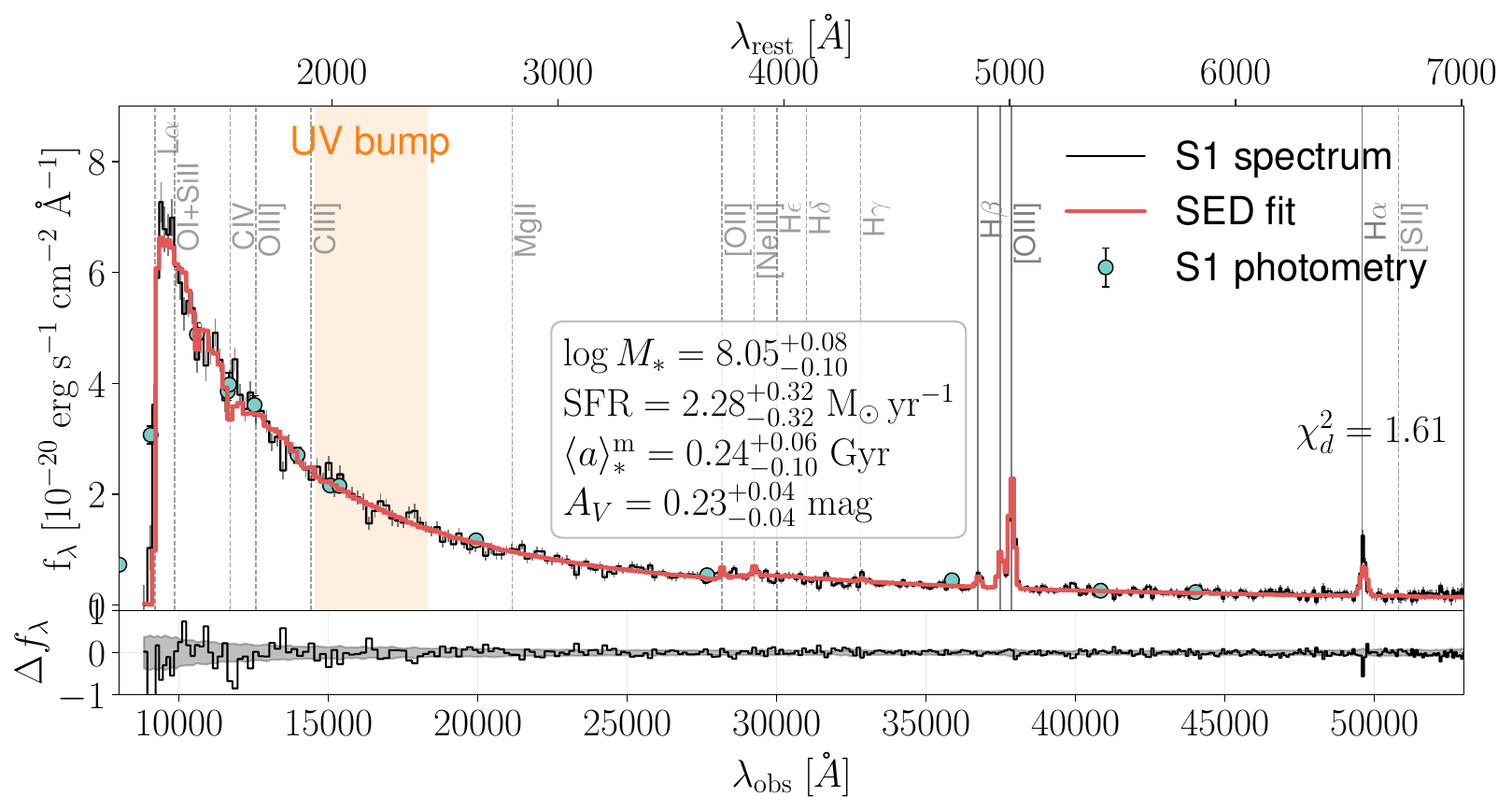}
    \includegraphics[width = 0.49\textwidth]{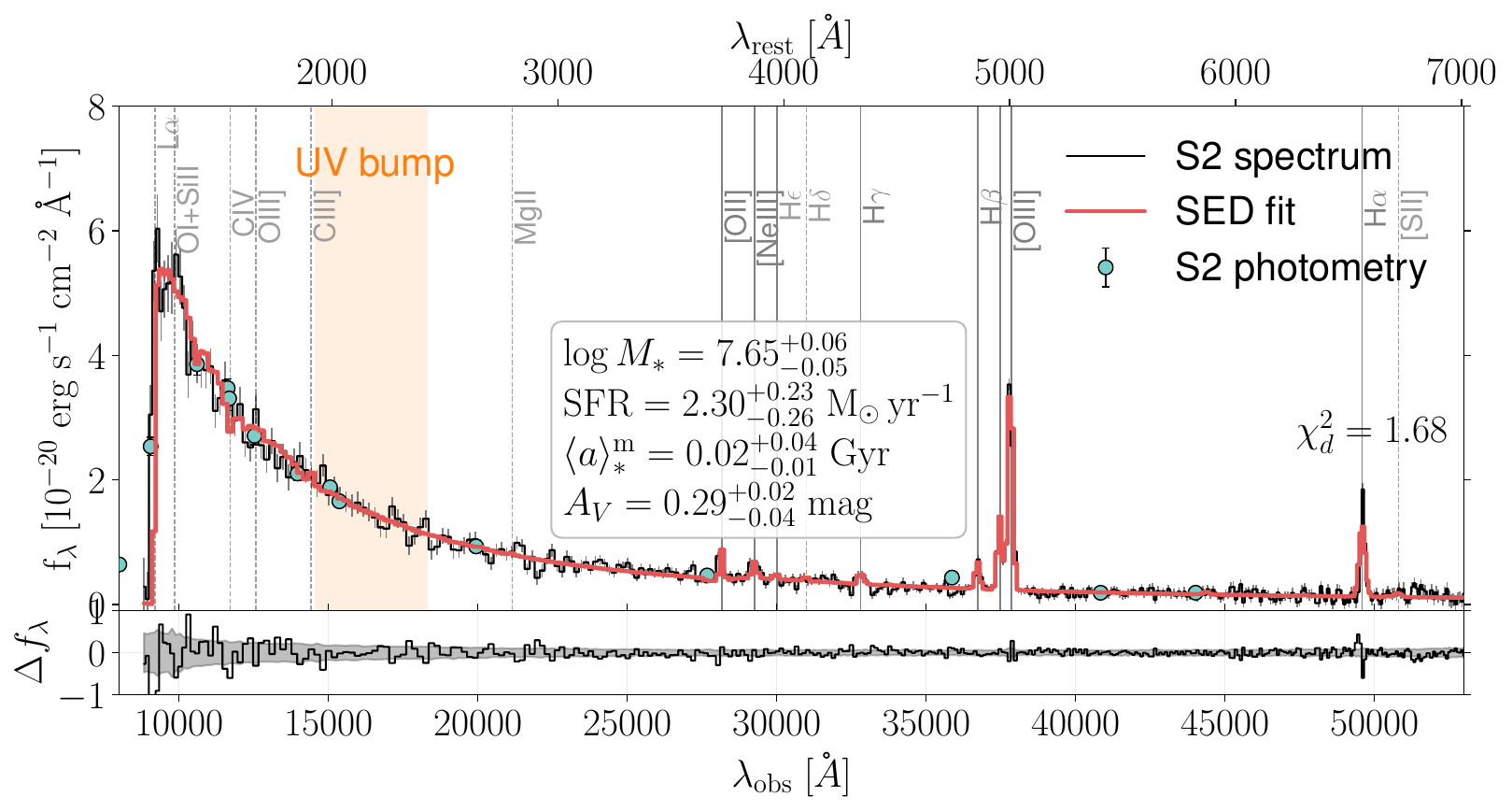} 
    \includegraphics[width = 0.49\textwidth]{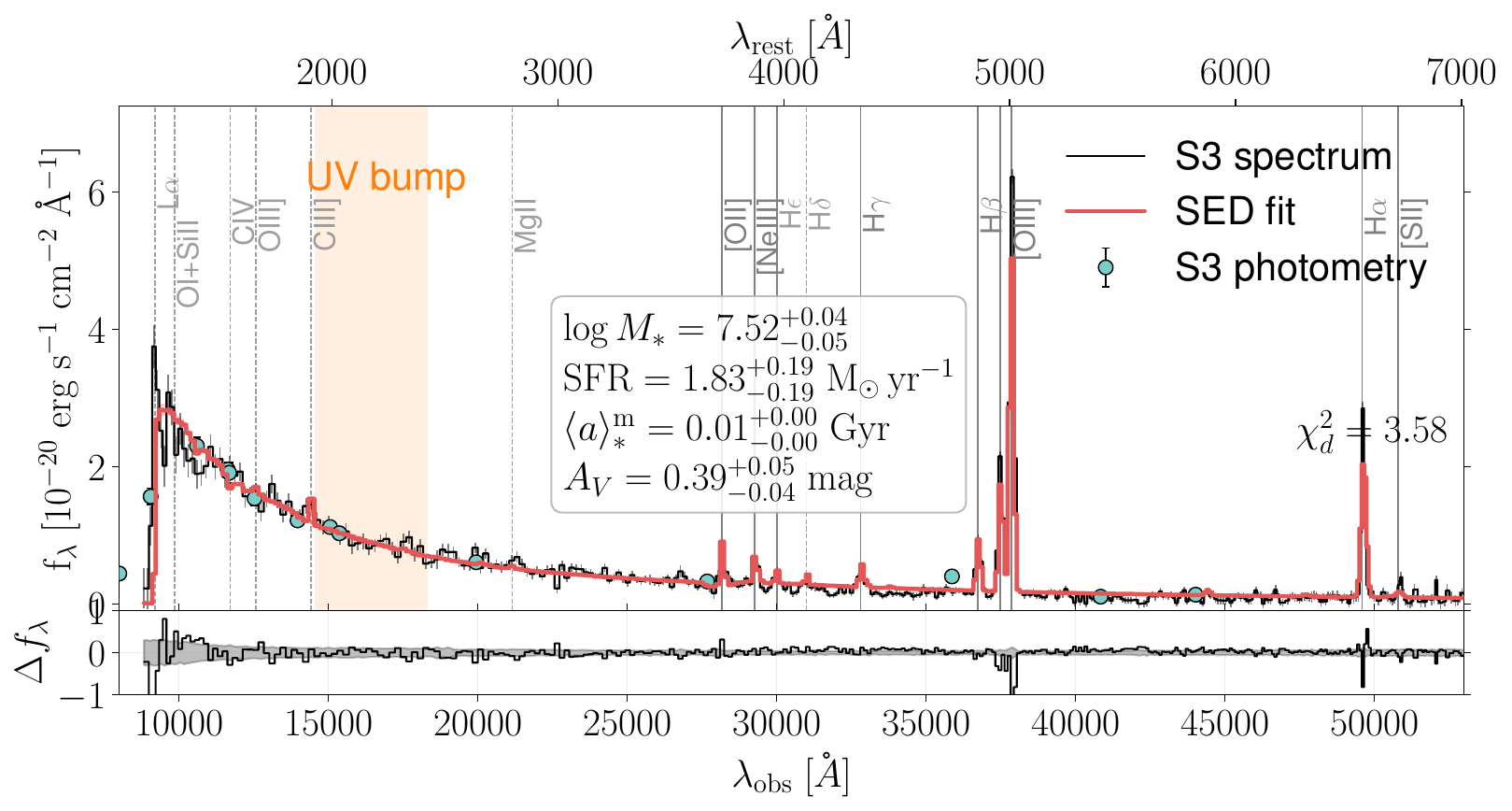}
\caption{NIRCam imaging, NIRSpec/PRISM spectroscopy, and the spectro-photometric fit for the bright LAE HCM 6A at $z = 6.5676$. Top left: RGB composite image of the system constructed from {\it JWST}/NIRCam 20 mas imaging by assigning F200W to the red channel, F150W to the green channel, and F115W to the blue channel. Middle left: NIRCam F115W image of the system. The three adjacent NIRSpec slits (NIRSpec IDs: 2101494, 2190001, and 2190002; S1, S2, and S3 hereafter) are overlaid on the RGB and F115W images targeting the three main clumps C1-C3. Clumps C1 and C3 exhibit internal substructure and are composed of mini-clumps (C1a-C1b and C3a–C3e, respectively), ordered from brightest to faintest in the RGB image. Top right and bottom panels: Slit-level photometry (teal circles), photometry-rescaled NIRSpec PRISM slit spectroscopy (black line), and the slit-level spectro-photometric fit using customized \texttt{BAGPIPES} (red line). Solid and dashed vertical lines indicate detected and undetected UV–optical emission lines, respectively. 
The vertical orange strip indicates the wavelength range of the UV bump absorption feature. The reduced chi-square, $\chi^2_\nu$, for each fit is shown in the top right. Inset panels: Basic galaxy properties derived from the SED fitting. 
}
    \label{fig:nircam_nirspec}
\end{figure*}

 Owing to its bright Ly$\alpha$ emission and multi-component morphology, HCM 6A was selected as a high-priority target for NIRSpec follow-up, with three adjacent slits (NIRSpec IDs: 2101494, 2190001, and 2190002; S1, S2, S3, hereafter) approximately centered on its three main clumps (C1, C2, and C3, hereafter; see Fig. \ref{fig:nircam_nirspec}).

\subsection{Previous multi-wavelength observations of HCM 6A} \label{data_LAE}

HCM 6A is a well-known, clumpy, and luminous LAE at $z = 6.5676$ behind the lensing cluster Abell 370 (\citealp{2002ApJ...568L..75H, 2005ApJ...635L...5C, 2007A&A...475..513B, 2013ApJ...771L..20K, 2020ApJ...896..156F}). This lensed source (with lensing magnification $\mu \approx 8.3-9.1$) was first identified as an LAE candidate using Keck/LRIS imaging and subsequently spectroscopically confirmed with LRIS (\citealp{2002ApJ...568L..75H}). Its bright Ly$\alpha$ emission shows a strongly asymmetric profile with a steep blue cutoff, consistent with scattering in a neutral medium. 
Star formation rate estimates span $\sim 2$–$40\ M_\odot,\rm yr^{-1}$, depending on whether they are derived from Ly$\alpha$ or UV continuum luminosity ($L_{\rm UV} = 2 \times 10^{29}\ \rm erg\ s^{-1}\ Hz^{-1}$). Next, HCM 6A has also been spectroscopically confirmed with Keck/DEIMOS (\citealp{2020ApJ...896..156F}). Parameters derived from the Ly$\alpha$ emission line measurements from \cite{2002ApJ...568L..75H} and \cite{2020ApJ...896..156F} are reported in Table \ref{Lya_params}.

\begin{table}[h!]
\centering
 \caption{Parameters derived from the Ly$\alpha$ emission line measurements from the literature. 
 \label{Lya_params}
 }
\begin{tabular}{lcc}
 \hline \hline
  \noalign{\smallskip}
  Reference & 1 & 2   \\
   \noalign{\smallskip}
    \hline 
   \noalign{\smallskip}
$z_{\rm Ly\alpha}$ &  $\approx 6.56$ & $6.572 \pm 0.001$ \\
\noalign{\smallskip}
$\rm EW_{\rm Ly\alpha} [\AA]$ & $\approx 25.1$ & $79.5 \pm 6.5$ \\
\noalign{\smallskip}
$f_{\rm{Ly\alpha}}  [\times 10^{-18} \rm erg\ cm^{-2}\ s^{-1}]$ & $\approx 6.5$ & $6.3 \pm 0.4$ \\ 
\noalign{\smallskip}
 \hline
\end{tabular}
\tablefoot{The listed parameters include spectroscopic redshift $z_{\rm Ly\alpha}$, rest-frame equivalent width EW, and delensed flux $f_{\rm{Ly\alpha}}$.}
\tablebib{
(1)~\citet{2002ApJ...568L..75H}; (2)~ \citet{2020ApJ...896..156F}}
\end{table}

HCM 6A has also been detected with {\it Spitzer}/IRAC at 3.6 and 4.5 $\mu m$ tracing rest-frame optical emission (\citealp{2005ApJ...635L...5C}). From the H$\alpha$ flux, an SFR estimate of $> 140 \ M_\odot,\rm yr^{-1}$ is inferred - higher than estimates based on Ly$\alpha$ and UV, consistent with the dust extinction of $A_V \sim 1$. They estimated a stellar mass of $M_* = 8.4 \times 10^8 \ M_\odot$.

Millimeter observations of HCM 6A with the IRAM 30M and NOEMA resulted in a non-detection of the \ion{C}{II} $158 \ \mu m$ and the continuum emission (\citealp{2007A&A...475..513B, 2013ApJ...771L..20K}). The estimated upper limit on the dust mass is $M_{\rm{dust}} < 5.3 \times 10^7 \ M_{\odot}$ (\citealp{2007A&A...475..513B}). HCM 6A has also been observed with ALMA Band 6 (2018.1.00035.L; PI Kotaro Kohno; \citealp{2024ApJS..275...36F}), at the edge of the map where the sensitivity is relatively low, resulting in a non-detection, with a $3\sigma$ upper limit of $ \sim 0.3 \ \rm{mJy}$, corresponding to a de-lensed dust mass upper limit of $M_{\rm{dust}} \lesssim 5 \times 10^6 M_{\odot}$ (following the methodology of \citealp{2000MNRAS.315..115D, 2012MNRAS.425.3094C, 2014PhR...541...45C}). 
Finally, {\it Chandra} did not detect an X-ray counterpart to HCM 6A (\citealp{2000ApJ...543L.119B}), suggesting the absence of a strong active galactic nucleus (AGN).

\section{Methodology} \label{Method}

\subsection{Ly$\alpha$ mapping}

We employ \texttt{SLEUTH} (\citealp{2024MNRAS.532..577E, 2025ApJ...991..188E}) to  the NIRISS WFSS spectrum with the F090WN filter (Sect. \ref{NIRISS_data}), to derive a high-quality Ly$\alpha$ emission line map of the LAE. 
\texttt{SLEUTH} is a spatially resolved grism-modeling pipeline that produces clean emission-line maps, accounting for spatially varying stellar populations. \texttt{SLEUTH} segments each galaxy into small regions—grown from the brightest pixel through nearest-neighbor expansion, until a target S/N of 20 is achieved, yielding roughly 30 regions across the source. For each region, \texttt{SLEUTH} forward-models spectral templates, including emission lines, and fits them to the observed grism spectra. The method explicitly captures spatial variations in stellar populations and line strengths while minimizing continuum leakage. 

\begin{figure}[h]
    \centering
    \includegraphics[width=\hsize]{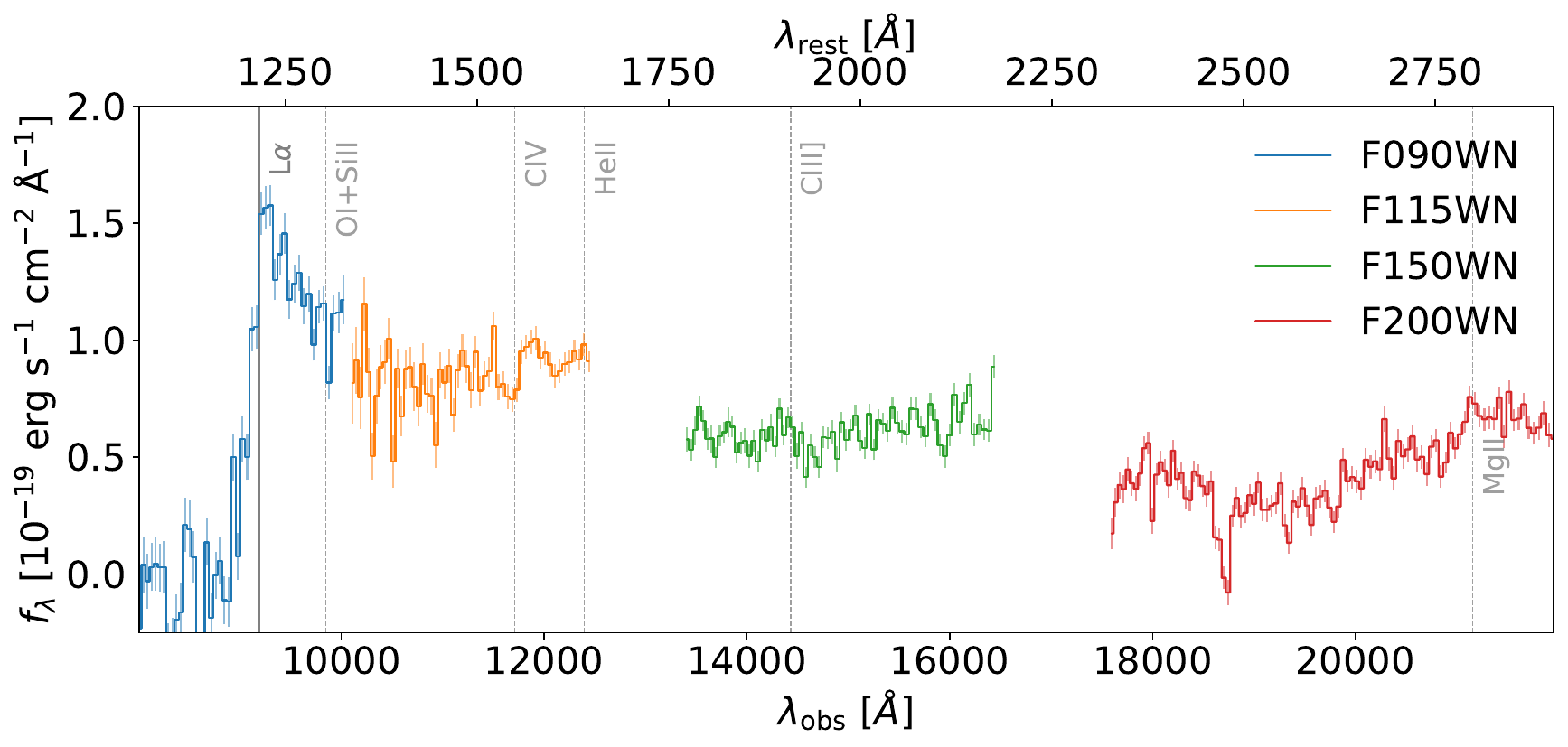}
\caption{{NIRISS 1D slitless spectrum of HCM 6A. The spectrum combines extractions from the NIRISS GR150C and GR150R grisms crossed with the F090WN, F115WN, F150WN, and F200WN filters. The spectrum shows a clear detection of the Ly$\alpha$ emission line at $\lambda_{\rm obs} \approx 9200 \AA$, consistent with $z \approx 6.6$ (solid vertical line). Dashed vertical lines indicate the expected positions of selected undetected emission lines. 
}
    \label{fig:lya_spec}}
\end{figure}

The extracted Ly$\alpha$ emission map was PSF-matched to the NIRCam/F115W resolution and resampled to a pixel scale of 0.04". At the source's redshift, this corresponds to $\approx75$ pc per pixel for an isotropic magnification of $\sim 8.3-9.1$. However, the true source-plane resolution is anisotropic: $\approx 25$ pc in the tangentially magnified direction (with $\mu_{\rm{tan}} \approx 7.6-8.4$) and $\approx 180$ pc in the radial direction ($\mu_{\rm{rad}} \approx 1.1$).
To complement the Ly$\alpha$ map, we extracted a 1D NIRISS spectrum from the GR150C and GR150R grisms, crossed with the four wide-band filters (F090WN–F200WN). The Ly$\alpha$ emission line is clearly visible at $\lambda_{\rm obs} \approx 9200\AA$, fully consistent with a redshift of $z \approx 6.6$ (Fig. \ref{fig:lya_spec}).

\subsection{Gravitational lens modeling and size measurements}

The lens model was constrained using the parametric lens modelling code \texttt{Lenstool} \citep{Kneib96, 2007NJPh....9..447J}. 
We adopted the lens model parameterisation described in \cite{2024ApJ...973...77G}. We further improved the model by incorporating a newly identified and confirmed multiple-image system at $z = 5.97$, thereby constraining the model using a catalog of 119 multiple images. The updated lens model will be available on MAST\footnote{\url{https://archive.stsci.edu/hlsp/canucs}}.
The magnification of the LAE, derived from the Bayesian samples of this model, lies in the range $\mu \approx 8.3$–$9.1$ across the source (i.e., NIRSpec slits). Most of this magnification arises from tangential stretching along the east-west direction, with a tangential component of $\mu_{\rm tan} \approx 7.6$-$8.4$. 

We measure the source and individual clump sizes directly in the source plane using forward modeling of the observed photometry with \texttt{Lenstruction} (\citealt{yang20}). The galaxy is modeled with a multi-component profile consisting of a Sérsic function for the underlying stellar component and the largest clump, as well as Gaussian profiles for the remaining individual clumps. The intrinsic source model is then iteratively mapped to the image plane using the gravitational-lensing maps and convolved with the empirical PSF.
For the entire source, we obtain a source-plane half-light radius of $ 66.0 \pm 0.3$ milli-arcseconds (mas), corresponding to $\approx 360$ pc at $z = 6.5676$.  The largest clump (C1a) has a best-fit size of $6.39 \pm 0.2$ mas ($\approx 35$ pc). However, the quoted uncertainties are likely underestimated; the true statistical uncertainties are expected to be at the $\approx 10-20\%$ level, depending on the adopted model. All the remaining clumps are unresolved, with $2\sigma$ upper limits of $<2.8$ mas ($< 15$ pc) on their intrinsic sizes.

\subsection{SED fitting with customized \texttt{BAGPIPES}}

Building on previous works (\citealp{2023A&A...679A..12M, 2025NatAs...9..458M, 2025A&A...702A..33M}), we have established a robust SED fitting framework capable of simultaneously constraining galaxy physical properties and the shape of the dust attenuation curve.
We apply our customized \texttt{BAGPIPES} SED-fitting pipeline \citep{2023A&A...679A..12M} to perform joint spectro-photometric fits and infer global ($\sim$kpc), slit-level ($\sim0.1$ kpc), and pixel-level ($\approx25$ pc) properties of our LAE source at $z = 6.5676$, including stellar mass (${M_*}$), SFR (both corrected for the local lensing magnification $\mu$), mass-weighted stellar age (${\langle a \rangle}_*^{\rm{m}}$), ionization parameter ($\log{U}$), metallicity ($Z$), and the $V$-band attenuation ($A_V$), along with the parameters ($c_1-c_4$) of the dust attenuation model (\citealp{2008ApJ...685.1046L}).

We modeled galaxy spectra using a customized version of the \texttt{BAGPIPES} SED fitting code (\citealp{2018MNRAS.480.4379C}), which employs a Bayesian framework with the MultiNest sampler (\citealp{2019OJAp....2E..10F}). Stellar emission was generated with the SPS models of \cite{2016MNRAS.462.1415C}, while nebular continuum and line emission were taken from precomputed CLOUDY grids (\citealp{2017RMxAA..53..385F}), with varying stellar age, metallicity, and ionization parameter. Our fiducial star formation history adopts a non-parametric continuity prior (\citealp{2019ApJ...876....3L}), which provides greater flexibility and minimizes biases compared to simple parametric forms. 
Our customized \texttt{BAGPIPES} (\citealp{2023A&A...679A..12M}) includes a flexible dust attenuation law (\citealp{2008ApJ...685.1046L}), which captures the expected diversity of attenuation curves at high redshift (e.g., \citealp{2025NatAs...9..458M, 2025A&A...702A..33M}). This formalism allows both the slope and the UV bump strength to vary, enabling us to explore departures from fixed empirical dust laws. The priors and their limits for each parameter of the SED fitting model are given in Table \ref{params}.

\begin{table}[ht]
\centering
 \caption{SED fitting parameters and their priors.}
 \label{params}

\begin{tabular}{lcccccccc}
 \hline \hline
  \noalign{\smallskip}
  Parameter & Limits & Prior \\
   \noalign{\smallskip}
 \hline
 \noalign{\smallskip}
$z$ & $(z_{\rm{sp}}-0.01, z_{\rm{sp}}+0.01)$ & Uniform \\
\noalign{\smallskip}
 $\sigma_{\rm{v}} \ [{\rm{km \ s^{-1}}}]$ & $(1, 2000)$ & Logarithmic \\
 \noalign{\smallskip}
 $\log{M_*^{\rm{form}} \ [M_{\odot}]}$ & $(4, 13)$ & Uniform \\
 \noalign{\smallskip}
$Z\ [Z_{\odot}]$  & $(0.001, 2.5)$ & Logarithmic \\ %
 \noalign{\smallskip}
 $\Delta \log(\rm{SFR})_i$ & $(-10, 10)$ & Student's-t \\
 \noalign{\smallskip}
$\log{U}$ & $(-4, 0)$ & Uniform \\ %
 \noalign{\smallskip}
$A_V \ [{\rm{mag}}]$ & $(0, 8)$ & Uniform \\ %
 \noalign{\smallskip}
 $c_1$ & $(0, 75)$ & Uniform  \\ %
 \noalign{\smallskip}
  $c_2$ & $(2, 20)$ & Uniform \\ %
 \noalign{\smallskip}
  $c_3$ & $(-2, 75)$ & Uniform \\ %
 \noalign{\smallskip}
  $c_4$ & $(-0.005, 0.1)$ & Uniform \\ %
 \noalign{\smallskip}
\hline
\end{tabular}

\tablefoot{The listed parameters include the redshift ($z$), velocity dispersion ($\sigma_{\rm{v}}$), the total stellar mass formed at the time of observation ($\log{M_*^{\rm{form}}}$), metallicity ($Z$), the logarithmic SFR ratio between adjacent time bins ($\Delta \log(\rm{SFR})_i$), ionization parameter ($\log{U}$), the $V$-band attenuation ($A_V$), and the the dust attenuation model parameters ($c_1-c_4$). $z_{\rm{sp}} = 6.5676$ is the spectroscopic redshift of the source from DAWN (\citealp{Brammer_msaexp_NIRSpec_analyis_2022, 2022zndo...7299500B}).}
\end{table}

We note that, although the spectroscopic redshift is securely determined from multiple emission lines, we allowed the redshift parameter in \texttt{BAGPIPES} to vary within a narrow prior of $z_{\mathrm{spec}}\pm0.01$. This was not intended to re-determine the redshift, but to allow limited flexibility to account for small mismatches between the low-resolution PRISM spectra and the model wavelength sampling. Fixing the redshift to $z_{\mathrm{spec}}$ yields physical parameters fully consistent within the quoted uncertainties.

 The analytical form of the dust attenuation model is given by:

\begin{equation}
\begin{split}   
A_{\lambda}/A_V &= \frac{c_1}{(\lambda/0.08)^{c_2}+(0.08/\lambda)^{c_2} +c_3} \\ 
& +  \frac{233[1-c1/(6.88^{c_2}+0.145^{c_2}+c_3)-c_4/4.60]}{(\lambda/0.046)^2+(0.046/\lambda)^2+90} \\ 
& + \frac{c_4}{(\lambda/0.2175)^2+(0.2175/\lambda)^2-1.95},
\end{split}
\end{equation}%
where $c_1-c_4$ are dimensionless parameters and $\lambda$ is the wavelength in $\mu m$. The three Drude-like components respectively model the far-ultraviolet rise, the optical–near-infrared attenuation, and the $\sim 2175 \AA$ bump.

For consistency with previous works (\citealp{2025NatAs...9..458M, 2025A&A...702A..33M}) and to facilitate comparison with literature results, we adopt the attenuation curve parameterization of \cite{2020ARA&A..58..529S}, which describes each curve in terms of two key quantities: the UV–optical slope ($S$) and the UV bump strength ($B$) at $\sim 2175 \AA$. The slope, $S$ is defined as $S = A_{1500}/A_V$, where $A_{1500}$ and $A_V$ are attenuation at $\lambda = 1500$ \r{A} and $V$-band, respectively. The UV bump strength, $B$ is defined as $B = A_{\rm{bump}}/A_{2175}$, where $A_{\rm{bump}}$ denotes the excess attenuation above the underlying continuum at $\lambda = 2175$ \r{A}, and $A_{2175}$ is the total attenuation at $\lambda = 2175$ \r{A}.

To derive the global physical properties of the galaxy on $\sim$ kpc scale, we performed SED fitting using the total {\it HST}+NIRCam photometry of the entire system. We used the catalog \texttt{COLOR03} aperture fluxes and scaled them to approximate total fluxes using the \texttt{KRON}/\texttt{COLOR03} flux ratio measured in the F277W band. This scaling ensures consistent total-flux normalization across all bands, while preserving the high S/N colors required for reliable global SED constraints.

Next, to derive spatially resolved constraints on key physical properties, we carried out spectro-photometric SED fitting by combining the NIRSpec/PRISM spectra (reduced with \texttt{msaexp} v4; \citealp{msaexp}) with matching {\it HST} + NIRCam photometry extracted over the same aperture. NIRSpec slit has a width of $0.2''$, corresponding to $\sim0.37$ kpc in the source plane, or $\sim0.13$ kpc along the tangential direction (given a magnification $\mu \approx 8.3$–9.1).
Each spectrum was rescaled with corresponding photometry to correct for slit losses and ensure consistent flux calibration. In this procedure, the photometric bands with S/N $< 2$ and rest-frame wavelengths $\lambda_{\rm rest} \lesssim 1300 \text{\AA}$ were excluded to avoid contamination from Ly$\alpha$ emission and the damping wing region.
Synthetic photometry was generated by convolving the spectra with the filter transmission curves using Dense Basis (\citealp{2019ApJ...879..116I}). The ratio of observed to synthetic photometry was fit with a first-degree Chebyshev polynomial to derive a linear spectral correction, preventing the artificial introduction of a UV bump around $\lambda \sim 2175 \text{\AA}$.

The rescaled spectra were then masked in regions of negative flux and blueward of Ly$\alpha$ to remove noisy slit-edge data and possible IGM absorption.
The resulting photometry, spectroscopy, and best-fit SED models for the three slits are shown in Fig.~\ref{fig:nircam_nirspec} (top right and bottom panels).

Finally, we performed pixel-level SED fitting using 19-band {\it HST} + NIRCam imaging to derive spatially resolved properties on a pixel scale of $0.04''$, corresponding to $\approx25$ pc in the source plane, owing to lensing. 
We restricted the analysis to pixels with ${\rm S/N} > 10$ in the $\chi_{\mathrm{mean}}$ detection image, which combines multiple bands weighted by their depth to optimize source detection across the field (see \citealp{2026ApJS..282....3S} for details). 
In addition, for this part of the analysis, we adopted a physically motivated logarithmic metallicity prior of $Z/Z_\odot \in [0.001, 1]$, given the limited sensitivity of low S/N, broadband photometry to infer $Z$, ensuring physically plausible solutions.  

Together, these complementary analyses provide a robust cross-validation of the LAE’s physical properties inferred from SED fitting across multiple spatial scales - global ($\sim$kpc), slit-level ($\sim0.13$ kpc), and pixel-level ($\sim25$ pc).

\subsection{$A_V$ from Balmer decrement and UV slope $\beta$}

We derived the UV slope ($\beta$) values using the \texttt{msaexp} package (\citealp{Brammer_msaexp_NIRSpec_analyis_2022, 2022zndo...7299500B, 2024Sci...384..890H, 2025A&A...697A.189D}) by fitting a power-law to the photometry corrected NIRSpec/PRISM spectra in the rest-frame UV, specifically within the two wavelength windows: $1400-1860 \AA$ and $1955-2580 \AA$, selected to avoid contamination from the \ion{C}{III]} emission line.
We also estimated the dust attenuation $A_V^{\rm B}$ from the Balmer decrement, defined as the ratio of the observed to intrinsic H$\alpha$/H$\beta$ line fluxes. The observed H$\alpha$/H$\beta$ ratios: ${\mathrm{H}\alpha/\mathrm{H}\beta}^{\rm{obs}} =  2.9 \pm 0.6$ for S1, ${\mathrm{H}\alpha/\mathrm{H}\beta}^{\rm{obs}} =  2.3 \pm 0.4$ for S2, and ${\mathrm{H}\alpha/\mathrm{H}\beta}^{\rm{obs}} =  2.3 \pm 0.2$ for S3; were measured from our emission line fits following the methodology of \cite{2025NatCo..16.9830T}. The intrinsic ratio (${\mathrm{H}\alpha/\mathrm{H}\beta}^{\rm{int}} \simeq 2.78$) was derived using the \texttt{PyNeb} software package (\citealp{2015A&A...573A..42L}), assuming an electron temperature of $T_{e^-} \simeq 15 000$ K and an electron density of $n_{e^-} = 1000\ \mathrm{cm^{-3}}$ (\citealp{Strom_2017, 2024ApJ...962...24S}).

\section{Results and discussion}\label{Results}

\subsection{Global properties}

\begin{table}[ht]
\centering
 \caption{{Constrained global properties of HCM 6A. }}
 \label{global_params}
\begin{tabular}{lc}
 \hline \hline
 \noalign{\smallskip}
 \multicolumn{2}{c}{SED fitting}\\
 \noalign{\smallskip}
\hline
 \noalign{\smallskip}
$\log{M_*\ [M_{\odot}]}$ & $8.3 - 8.4$ \\
 \noalign{\smallskip}
$\rm{SFR\ [M_{\odot} \ yr^{-1}]}$ & $2.4 - 2.7$ \\
 \noalign{\smallskip}
$\rm{sSFR\ [Gyr^{-1}]}$ & $10.7_{-1.4}^{+0.5}$ \\
 \noalign{\smallskip}
${\langle a \rangle}_*^{\mathrm{m}}/ \rm{Myr}$ & $35_{-23}^{+61} $ \\
 \noalign{\smallskip}
$\log{U}$ & $-1.6 \pm 0.2$ \\ %
 \noalign{\smallskip}
$Z \ [Z_{\odot}]$  & $0.25_{-0.05}^{+0.05} $ \\ %
 \noalign{\smallskip}
$A_V\ [{\rm{mag}}]$ & $0.19_{-0.06}^{+0.03}$ \\ %
 \noalign{\smallskip}
 $S$ & $2.32 \pm 0.03$  \\ %
 \noalign{\smallskip}
  $B$ & $0.05 \pm 0.08$ \\ %
 \noalign{\smallskip}
 \hline
  \noalign{\smallskip}
 \multicolumn{2}{c}{UV flux measurements} \\
 \noalign{\smallskip}
\hline
 \noalign{\smallskip}
  $M_{\rm{UV}}$ [mag] & $-19.8 \pm 0.1$ \\ %
 \noalign{\smallskip}
  $L_{\nu} \ [10^{28}  \rm{erg \ s^{-1} \ Hz^{-1}]}$ & $3.6 \pm 0.2$ \\ %
 \noalign{\smallskip}

\hline
\end{tabular}
\tablefoot{$\log M_*$, SFR, $M_{\rm{UV}}$, and $L_{\nu}$ are corrected for the lensing magnification.}
\end{table}

We derived the global physical properties of the galaxy by fitting the total combined {\it HST}+NIRCam photometry, obtaining SFR and $M_*$ ($\mu$-corrected), specific SFR  (sSFR), ${\langle a \rangle}_*^{\rm{m}}$, $\log U$, $Z$, $A_V$, along with attenuation curve parameters $S$ and $B$ (following the parametrization of \citealp{2020ARA&A..58..529S}). 
Additionally, we estimated the intrinsic UV luminosity of the system, using the total flux in the F115W band, which closely probes the rest-frame $\sim 1500 \AA$ continuum at $z = 6.5676$. The measured flux of $f_{\nu} =  0.446 \pm 0.019 \ \rm{\mu  Jy}$ corresponds to an observed absolute magnitude of $M_{\rm{UV}}^{\rm{obs}}  = -22.1 \pm 0.1$. After correcting for the lensing magnification ($\mu = 8.3-91$, the intrinsic UV luminosity of the galaxy is $M_{\rm{UV}}  = -19.8 \pm 0.1$, consistent with the estimate reported by \cite{2020ApJ...896..156F}. 
The intrinsic UV magnitude corresponds to a monochromatic luminosity of $L_{\nu} = 3.6 \pm 0.2 \times 10^{28}  \rm{erg \ s^{-1} \ Hz^{-1}}$, a factor $\sim 5$ lower than the estimate by \cite{2002ApJ...568L..75H}. This luminosity is typical of moderately luminous galaxies at the EoR (e.g., \citealp{2010ApJ...724.1524O, 2024A&A...682A..40W, 2025MNRAS.536...27W, 2025ApJ...988...26W}).
Global physical properties of HCM 6A are reported in Table~\ref{global_params}.

\subsection{Slit-level properties}

\begin{figure*}[h]
    \centering
    \includegraphics[width = \textwidth]{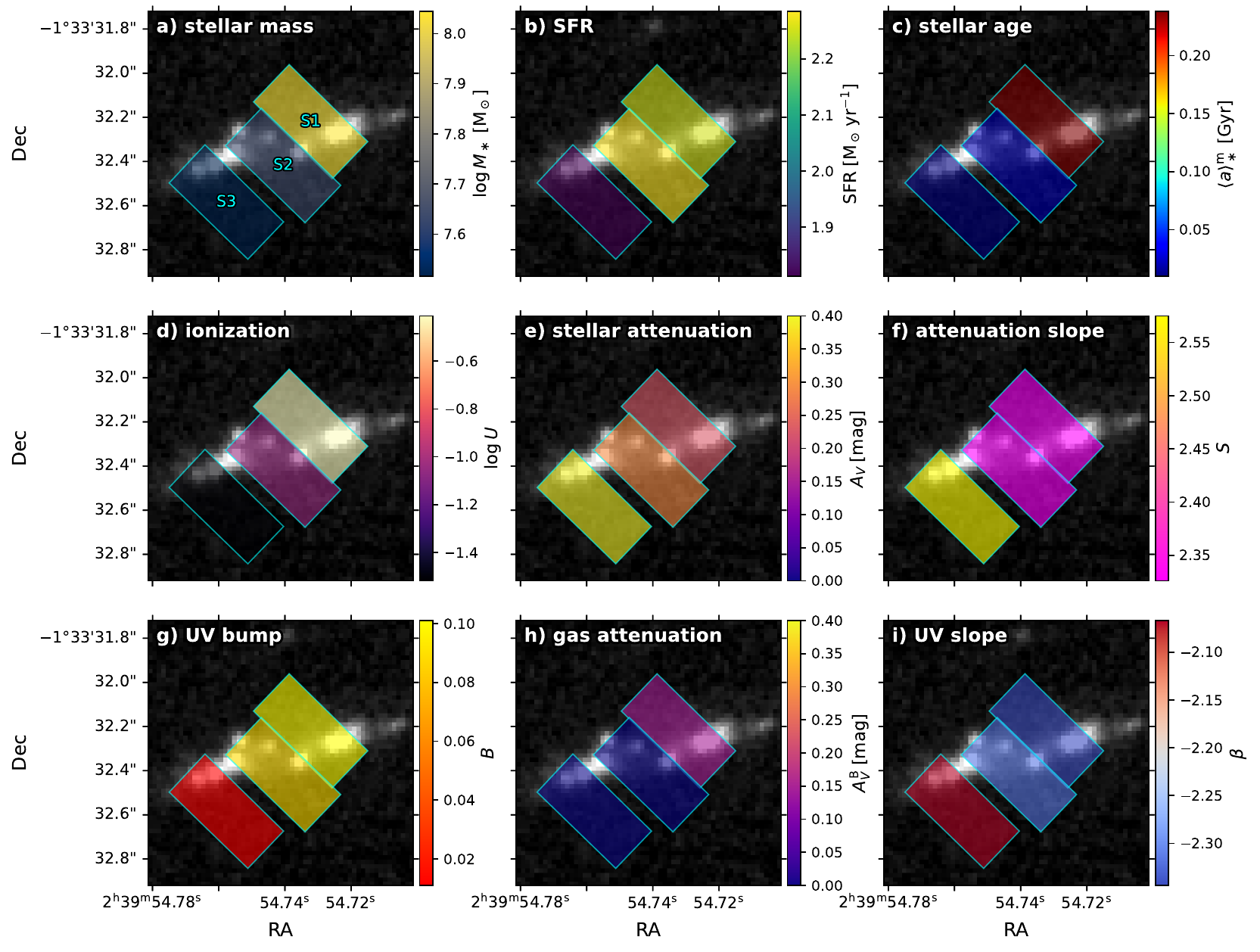}
    \caption{Slit-level, spatially resolved physical properties of the target system:
    stellar mass ($\log M_*$; a), star formation rate (SFR; b),
    mass-weighted stellar age (${\langle a \rangle}_*^{\mathrm{m}}$; c),
    ionization parameter ($\log U$; d), $V$-band attenuation ($A_V$; e),
    slope ($S$; f) and UV bump ($B$; g) of the dust attenuation curve,
    attenuation derived from the Balmer decrement ($A_V^{\mathrm{B}}$; h),
    and UV continuum slope ($\beta$; i). Both $\log M_*$ and SFR are corrected for the lensing magnification. The NIRCam 20mas F115W map of the system is shown as the background image.}
    \label{fig:slit_prop}
\end{figure*}

We performed joint spectro-photometric fitting, combining {\it HST} and NIRCam slit-level ($\sim0.37$ kpc in the source plane, or $\sim0.13$ kpc tangentially) photometry with photometry-rescaled NIRSpec PRISM spectra obtained in three adjacent slits (S1-S3) roughly centered on the three main clumps (C1-C3; Fig. \ref{fig:nircam_nirspec}). This {\it mini-IFU} configuration provides slit-level, spatially resolved key physical properties from the SED fits: SFR and $M_*$ ($\mu$-corrected), ${\langle a \rangle}_*^{\rm{m}}$, $\log U$, $Z$, $A_V$, along with attenuation curve parameters.

The delensed stellar masses are $\log M_* = 8.04 \pm 0.10$ for S1, $\log M_* = 7.75 \pm 0.03$ for S2, and $\log M_* = 7.51 \pm 0.02$ for S3. The combined intrinsic stellar mass of the three slit regions, $\log M_* = 8.30 \pm 0.06$, is in excellent agreement with the total unlensed stellar mass inferred from the integrated {\it HST}+NIRCam photometry ($\log M_* \simeq 8.3$–$8.4$;  Table~\ref{global_params}).

In addition, we independently estimated the slit-level dust attenuation from the Balmer decrement ($A_V^{\rm B}$) derived from the H$\alpha$/H$\beta$ ratio. We measured the UV continuum slope $\beta$ directly from the photometry-rescaled NIRSpec spectra. These complementary estimates allow us to cross-check and compare the attenuation inferred from the SED fitting ($A_V$), with that derived from nebular emission lines ($A_V^{\rm B}$), and from the stellar continuum ($\beta$). Fig. \ref{fig:slit_prop} depicts slit-level, spatially resolved stellar, nebular, and dust properties across our target galaxy. The metallicity map is omitted, as SED fitting may not provide reliable constraints. However, the fitted values span $Z \sim 0.2-0.7 \ Z\odot$ from the young, low-mass region (S3) to the more evolved one (S1).

The $A_V$ estimates from SED fitting indicate that all three slit regions are moderately dusty ($A_V \sim 0.2$–$0.4$), with S3 being the most attenuated ($A_V \sim 0.39 \pm 0.05$; Fig. \ref{fig:slit_prop}, panel e).
In contrast, attenuation derived from the Balmer decrement ($A_V^{\rm B}$) suggests a possible mild attenuation in S1 ($A_V^{\rm B} = 0.15 \pm 0.70$) and negligible attenuation in the remaining two regions ($A_V^{\rm B} \sim 0$; Fig. \ref{fig:slit_prop}, panel h), although the associated uncertainties are large. The attenuation derived from Balmer line ratios in S2 and S3 is slightly negative ($A_V^{\rm B} \lesssim 0$), which is unphysical. Such results are occasionally observed, particularly in low-mass, high-redshift galaxies (e.g., \citealp{2023ApJ...949L..11M, 2023ApJ...954..157S}). They are likely due to measurement uncertainties or deviations in gas temperature and density from the standard Case B recombination assumptions (\citealp{2006agna.book.....O}).  

While S1 appears moderately dusty across all attenuation indicators, the remaining two regions show a clear tension between the stellar and nebular dust probes. The discrepancy between the $V$-band attenuation derived from the SED modeling ($A_V$) and the one inferred from Balmer line ratios ($A_V^{\rm B}$) - most notably in S3 - may point to a complex geometry between stars, gas, and dust, where the stellar and nebular components are subject to different levels of dust obscuration (e.g., \citealp{2008ApJ...678..655F}). 
Since $A_V$ primarily traces attenuation of the UV–optical continuum, it is intrinsically more sensitive to stellar than to nebular line emission. This interpretation is also supported by the UV continuum slope, $\beta$, which also traces dust attenuation of the stellar populations. S3 exhibits the reddest UV slope ($\beta = -2.1 \pm 0.1$), corresponding to moderate stellar obscuration ($A_V \gtrsim 0.15$; assuming the \citealp{1999ApJ...521...64M} relation), while S1 and S2 show bluer slopes ($\beta \sim -2.3$; Fig. \ref{fig:slit_prop}, panel i; Fig. \ref{fig:three_spectra}), consistent with negligible stellar attenuation ($A_V \gtrsim 0.0$). These results reinforce the interpretation that S3 experiences the highest stellar attenuation among the three regions.

In contrast, the Balmer-derived attenuation, $A_V^{\rm B}$ - which primarily traces dust affecting nebular emission lines - indicates a weak attenuation in S1, with $A_V^{\rm B} \sim 0.15$ (although with large uncertainty) and no significant attenuation of the ionized gas in S2 and S3, with $A_V^{\rm B} \sim 0$. 
This interpretation is further supported by the strength of the UV and optical emission lines in the NIRSpec PRISM spectra across the three clumps (Fig. \ref{fig:nircam_nirspec}, Fig. \ref{fig:three_spectra}). Notably, S3 exhibits the brightest nebular emission lines (Fig. \ref{fig:nircam_nirspec}; bottom right panel, Fig. \ref{three_spectra}, inset panels). The potential presence of Ly$\alpha$, which is particularly sensitive to resonant scattering and dust absorption (e.g., \citealp{2014PASA...31...40D}), reinforces the conclusion that the nebular regions in S3 experience little to no dust attenuation. 

We note that a tentative detection ($\rm{S/N \sim 4}$) of the Ly$\alpha$ emission line is seen in the v3 NIRSpec reduction of the S3 region, whereas it is not significantly detected in v4. These two reductions correspond to independent background subtraction strategies, and the difference likely reflects reduction-dependent background treatment (see Appendix \ref{Lya_flux}). Repeating the analysis with the v3 spectra yields results consistent within 1–2$\sigma$, including the $A_V^{\rm B}$ values inferred from the Balmer decrement.

This is consistent with a very young stellar population (${\langle a \rangle}_*^{\rm{m}} = 10 \pm 1 \ \rm{Myr}$; Fig. \ref{fig:slit_prop}, panel c) and high sSFR in S3 ($\rm{sSFR\ \simeq 55 \ Gyr^{-1}}$) suggesting that radiation-driven outflows from a recent starburst may have cleared out much of the dust from the ionized regions (e.g., \citealp{2023MNRAS.526.4801T, 2023MNRAS.522.3986F, 2024A&A...684A.207F, 2025MNRAS.544.4390N}). 
The extreme inferred sSFR in S3 places this region in a regime consistent with a super-Eddington starburst, in which feedback is expected to be dominated by radiation pressure. In this scenario, efficient dust clearing by the outflow is naturally expected, a condition also indirectly required by the detection of Ly$\alpha$ emission (\citealp{2024A&A...684A.207F, 2025MNRAS.544.4390N}). The properties of S3 therefore closely resemble those predicted by attenuation-free models for compact, rapidly growing systems, making it a compelling analog of the “super-early” blue star-forming galaxies inferred at $z \gtrsim 10$ (\citealp{2024A&A...684A.207F, 2025A&A...694A.286F, 2025MNRAS.544.4390N}).

\begin{figure*}[h]
    \centering
    \includegraphics[width = \textwidth]{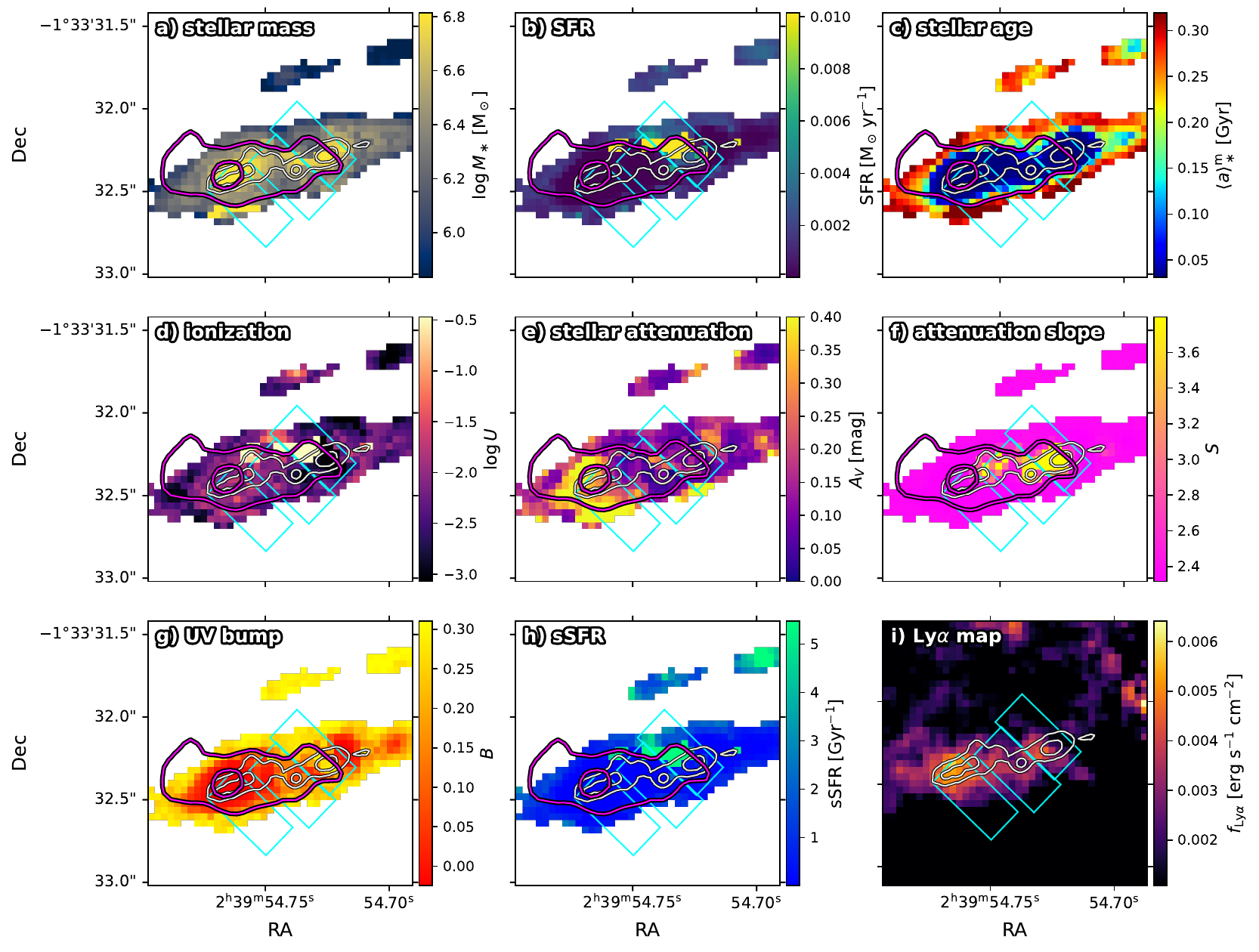}
    \caption{Pixel-level, spatially resolved physical properties of the target system:
    stellar mass ($\log M_*$; a), star formation rate (SFR; b),
    mass-weighted stellar age (${\langle a \rangle}_*^{\mathrm{m}}$; c),
    ionization parameter ($\log U$; d), $V$-band attenuation ($A_V$; e),
    slope ($S$; f) and UV bump ($B$; g) of the dust attenuation curve, sSFR (h), and
    Ly$\alpha$ flux ($f_{\mathrm{Ly\alpha}}$; i). The quantities $\log M_*$, SFR, and $f_{\mathrm{Ly\alpha}}$ are corrected for lensing magnification. The three adjacent NIRSpec slits are overlaid as cyan rectangles. The NIRCam RGB map is shown as white contours at the 5$\sigma$ and 10$\sigma$ levels, while the Ly$\alpha$ emission is traced by magenta contours at 1-2$\sigma$.
}
    \label{fig:pixel_prop}
\end{figure*}

The Ly$\alpha$-emitting clump (2$\sigma$ contours in Fig. \ref{fig:pixel_prop}) has an observed size of $0.08''\times 0.10''$, which, after correcting for lensing, corresponds to an effective source-plane radius of $R_{\rm eff} \sim 80$ pc—combined with the young stellar age ($ {\langle a \rangle}_*^{\rm{m}}\lesssim 10$ Myr), this yields a minimal characteristic clearing velocity of $v_{\rm{clear}} \gtrsim R_{\rm eff}/{\langle a \rangle}_*^{\rm{m}} \gtrsim 8~\rm km \ s^{-1}$. This quantity represents the minimum expansion rate required to remove or dilute dust and neutral gas over the extent of the Ly$\alpha$ cavity. This is consistent with modest ($\sim$ few–tens of $\mathrm{km,s^{-1}}$) local radiation- and stellar-feedback models, which predict low-velocity expansion of ionized cavities capable of opening the low-opacity channels necessary for Ly$\alpha$ escape (e.g., \citealp{2008MNRAS.391..457D, 2018MNRAS.475.4617K}).

In summary, the more massive, star-forming and older (Fig. \ref{fig:slit_prop}, panels a-c) region S1 appears moderately dusty across all attenuation indicators - suggesting a relatively uniform and more settled ISM geometry, expected for an evolved ($\sim240$ Myr) region. In contrast, the youngest region S3 shows clear tensions between the different attenuation diagnostics, while S2 represents an intermediate case. These tensions likely reflect the various physical components probed by each diagnostic — with $A_V$ and $\beta$ tracing stellar continuum attenuation, and $A_V^{\rm B}$ and emission line brightness tracing the nebular component — suggesting a complex dust geometry and more turbulent ISM in S3.
However, the limited spatial resolution of the slit-level spectro-photometric data ($\sim 0.1$ kpc) restricts our ability to resolve the small-scale structures and disentangle these components.

\subsection{Pixel-level properties}

To test the geometry-driven explanation for the observed dust attenuation discrepancies, we investigate the spatially resolved properties on a pixel scale by performing SED fitting of the 19-band photometry from {\it HST} and {\it JWST}. The strong lensing magnification ($\mu \approx 8.3$–$9.1$) allows us to probe extremely fine spatial scales, with the drizzled NIRCam pixel size of $0.04''$ corresponding to an average of $\approx$25 pc in the source plane. 
This resolution approaches the scale of individual giant {\Hii} regions and massive star-forming clumps (e.g., \citealp{2009A&A...507.1327H}), enabling a detailed view of the galaxy's internal structure and the interplay between dust, stars, and feedback. 

On identical physical scales, we also leverage NIRISS WFSS observations to generate a spatially resolved Ly$\alpha$ emission map (Fig. \ref{fig:pixel_prop}, panel i), providing additional insight into the spatial distribution of ionized gas and its relationship with dust attenuation on $\approx$25 pc scale. We measure a total integrated Ly$\alpha$ flux of $F_{\mathrm{Ly}\alpha} = (7.0 \pm 0.3) \times 10^{-17}\ \mathrm{erg\ s^{-1}\ cm^{-2}}$ across the entire galaxy (see Appendix \ref{Lya_flux}). After correcting for lensing magnification ($\mu \approx 8.3$–$9.1$), this corresponds to an intrinsic flux in the range of $F_{\mathrm{Ly}\alpha} = 8.2-9.0 \times 10^{-18}\ \mathrm{erg\ s^{-1}\ cm^{-2}}$, depending on the magnification ($\mu \approx 8.3-9.1$), consistent with previous measurements in the literature (\citealp{2002ApJ...568L..75H,  2020ApJ...896..156F}; see Table \ref{Lya_params}).

The spatially resolved pixel-level stellar, nebular, dust, and Ly$\alpha$ maps of our LAE are shown in Fig.~\ref{fig:pixel_prop}.
The properties $\log M_*$, SFR, and $f_{\mathrm{Ly\alpha}}$ are corrected for lensing magnification based on the (WCS-matched) magnification map, derived from the best-fit lens model, covering the region around the target galaxy.
The sum of the intrinsic stellar masses over all SED–fitted pixels within the source yields $\log M_\ast \simeq 9.1$, a factor of $\sim5$–$6$ higher than the total stellar mass inferred from the integrated photometry ($\log M_\ast = 8.3$–$8.4$) or from the sum of the three NIRSpec slit regions ($\log M_\ast \simeq 8.3$). Restricting the sum to only the pixels within the three slits reduces the discrepancy, but still produces $\log M_\ast \simeq 8.8$, i.e., a factor of $\sim3$ higher than the integrated and slit-based stellar mass estimates. This discrepancy is consistent with the well-known tendency of unresolved SED-based estimates to yield systematically lower stellar masses than resolved, pixel-by-pixel SED fitting, possibly due to the outshining effect (\citealp{2015MNRAS.452..235S, 2018MNRAS.476.1532S, 2023ApJ...948..126G, 2025MNRAS.542.2998H}).

Most of the pixel-level quantities show good overall agreement with the slit-level results (Fig. \ref{fig:slit_prop}), although the higher resolution of the pixel-based analysis reveals finer spatial variations.
The main exception is the mass-weighted stellar age, which does not show consistent trends between the two approaches (panels c of Figs. \ref{fig:slit_prop} and ~\ref{fig:pixel_prop}) - likely due to the limited ability of photometry-only SED fitting to constrain stellar ages (\citealp{2020MNRAS.495.2088C, 2024A&A...681A..94N, 2025A&A...695A..86N}), given the age–dust–metallicity degeneracies (see e.g., \citealp{2011Ap&SS.331....1W} for a review).
Overall, while pixel-level SED fitting enhances spatial resolution, the accuracy of parameters, such as stellar age, metallicity, dust attenuation, and UV slope ($\beta$), remains limited by the low S/N photometry and lack of spectroscopic coverage in individual pixels (\citealp{2024arXiv241114532S, 2025A&A...695A..86N}). In particular, Balmer-line attenuation ($A_V^{\rm B}$) requires robust spectroscopic detection of the H$\alpha$ and H$\beta$ lines. 

NIRISS Ly$\alpha$ emission line map shows that the Ly$\alpha$ emission is concentrated in clump C3, with its peak located around and slightly north of the C3b mini-clump (Fig.~\ref{fig:pixel_prop}, panel~i).
The $V$-band attenuation ($A_V$) also peaks around C3, consistent with the slit-level results (panels~e of Figs.~\ref{fig:slit_prop} and \ref{fig:pixel_prop}), seemingly at odds with expectations given the strong sensitivity of Ly$\alpha$ to scattering and absorption by dust (e.g., \citealt{1999ApJ...518..138H}).
However, the pixel-level maps provide a more detailed view of the spatial distribution: the Ly$\alpha$ emission is concentrated in the central region of C3, whereas the $A_V$ peaks in the regions surrounding the Ly$\alpha$ peak, towards the C3 outskirts ($A_V \sim 0.3-0.5$).
This spatial configuration suggests a complex ISM geometry in which the central region of C3 is largely ionized and relatively dust-free ($A_V \sim 0.1-0.3$), allowing Ly$\alpha$ photons to escape through cleared channels (e.g., \citealp{2008A&A...488..491A}) even through the partially neutral IGM (e.g., \citealp{2024A&A...684A.207F}). The inferred young age of this region (${\langle a \rangle}_*^{\rm{m}} \sim 10~\mathrm{Myr}$ in S3; Fig. \ref{fig:slit_prop}, panel c) indicates a recent star-formation episode that has likely expelled and/or destroyed much of the dust in the core through radiation-driven outflows, displacing it toward the periphery of the clump, possibly via radiation-driven outflows (e.g., \citealp{2024A&A...684A.207F, 2025MNRAS.544.4390N}).
We note, however, that we are probably observing a relatively dust-cleared sightline toward the C3 core; from other orientations, dust in the outskirts could obscure the central emission and suppress the observed Ly$\alpha$ escape.
These results also show that the presence of dust does not necessarily inhibit Ly$\alpha$ escape: instead, the detailed geometry of dust and stars-likely sculpted by stellar feedback-plays a determining role in regulating Ly$\alpha$ transmission.

\subsection{Attenuation curve properties}

\begin{figure}[h]
    \centering
    \includegraphics[width=\hsize]{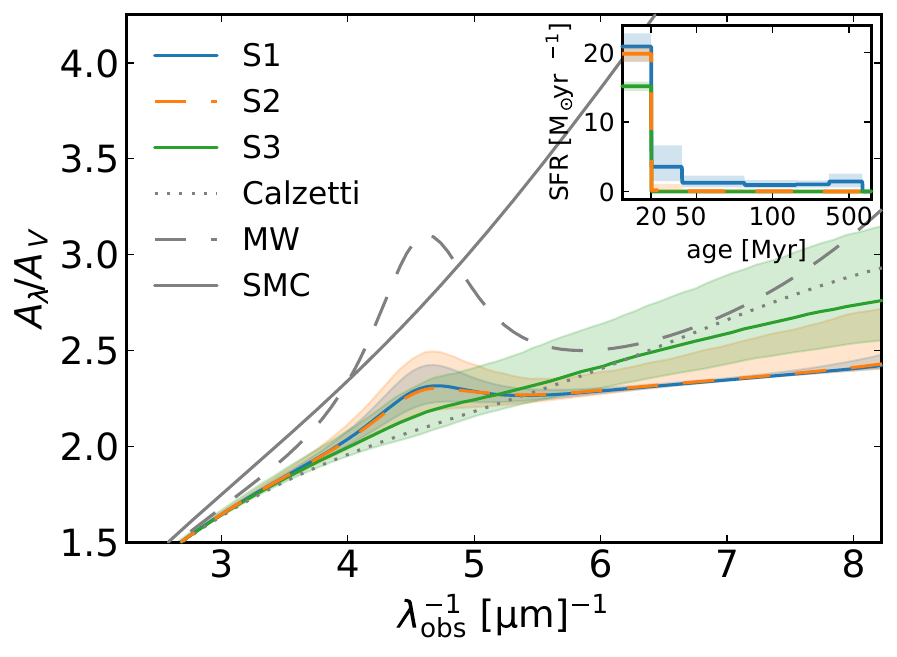}
\caption{{Dust attenuation curves of the three slit regions (S1–S3) at $z = 6.5676$ derived from the slit-level spectro-photometric fits. S1, S2, and S3 are shown as blue solid, orange dashed, and green solid lines, respectively. Standard Milky Way (MW), Calzetti, and Small Magellanic Cloud (SMC) attenuation curves are overplotted for comparison (gray dashed, dotted, and solid lines, respectively). Inset panel: corresponding star formation histories (SFHs) of the three slit regions.
}
    \label{fig:att_sfh}}
\end{figure}

In Fig. \ref{fig:att_sfh}, we show the attenuation curves reconstructed from the flexible dust model parameters ($c_1$–$c_4$; \citealp{2008ApJ...685.1046L}) implemented in our customized \texttt{BAGPIPES} framework (\citealp{2023A&A...679A..12M}).
From the global SED fit (Table \ref{global_params}) to the combined photometry and rescaled NIRSpec spectra of the three slit regions (Fig. \ref{fig:att_sfh}), the inferred attenuation curves are predominantly flat and largely featureless - i.e., "Calzetti-like" (\citealp{2000ApJ...533..682C}) - consistent with recent high-$z$ observations ($z \sim 6.3-11.5$; \citealp{2025NatAs...9..458M}) and predictions ($z \sim 6$; \citealp{2018ApJ...869...70N}) indicating dust dominated by large, supernova (SN)-produced grains with limited ISM reprocessing (e.g., \citealp{2025NatAs...9..458M, 2025ApJ...985L..21M}). 
Nevertheless, a measurable variation in slope and UV-bump strength can be observed across the clumps.

At the slit level, the attenuation curve slope ($S$) remains relatively constant ($S \sim 2.3$–$2.6$;  Fig. \ref{fig:slit_prop}, panel f) across the three clumps, varying by $ \sim 10\%$. Given this limited variation, we refrain from over-interpreting potential trends involving $S$. In contrast, the UV bump strength ($B$) varies more (Fig. \ref{fig:slit_prop}, panel g), ranging from being negligible in S3 to being more prominent in S1 and S2. Notably, $B$ increases with decreasing $A_V$ (Fig. \ref{fig:slit_prop}, panel e) and increasing age (Fig. \ref{fig:slit_prop}, panel c), consistent with the global trends observed in e.g., \cite{2020ApJ...888..108B, 2025A&A...702A..33M}.

We report a tentative ($\sim 2.6\sigma$) detection of a weak UV bump in the slit region S1 of the LAE, with the UV bump parameter $c_4 = 0.012 \pm 0.005$ ($B = 0.10 \pm 0.05$), corresponding to $\sim 23\%$ ($\sim 28\%$) of the Milky Way bump. 
To confirm the presence of the $2175 \AA$ UV bump in this source, we used $\chi^2$ statistics (e.g., \citealp{2016A&A...588A..19L}) to perform a nested model comparison between (i) the fiducial SED model (with a free UV bump amplitude $c_4$) and (ii) the same model without a bump (with $c_4 = 0$). The comparison yields only weak evidence for the presence of the feature in the spectral region around the bump ($P \simeq 0.25$), whereas the full-spectrum fit shows a significant improvement when including the bump component ($P < 0.05$).

At the pixel level, both the dust attenuation curve slope ($S$) and UV bump amplitude ($B$) exhibit a broader range of variation compared to the slit-level results (panels f–g of Figs.~\ref{fig:slit_prop} and \ref{fig:pixel_prop}). In addition, both $S$ and $B$ peak around the position of clump C1 (C1a), extending slightly toward C2, with a smaller peak around the C3e mini-clump (panels f–g of Fig.~\ref{fig:slit_prop}).
The SFR (averaged over 10 Myr), sSFR, and ionization parameter ($\log{U}$; Fig.~\ref{fig:slit_prop}, panels b, i, and d, respectively) also peak between and northward of C1 and C2,
with smaller peaks around C3d-e, potentially tracing merger-induced recent star formation. The resulting strong radiation fields in these regions may drive the reprocessing of dust grains and the production of smaller grains (including carbonaceous grains or PAHs), leading to steeper slopes and stronger UV bumps (\citealp{2023ApJ...951..100N, ormerod2025detection2175aauvbump}). 

This behavior contrasts with trends typically observed on global galaxy scales, where steeper slopes and stronger bumps are typically associated with lower sSFR and older stellar populations (e.g., \citealp{2013ApJ...775L..16K, 2020ApJ...888..108B, 2025A&A...702A..33M}). This discrepancy likely reflects the different drivers of sSFR on local versus global scales: globally, sSFR is driven mainly by the inverse of stellar age (e.g., \citealp{2024arXiv241014671L}) or stellar mass (by definition); locally, sSFR variations arise mostly from local variations in SFR at nearly fixed stellar mass within a single clump (Fig. \ref{fig:pixel_prop}). As a result, the pixel-level correlation between $S$, $B$, and SFR (sSFR) probably traces spatially-localized star formation and feedback, whereas the global anti-correlation with sSFR reflects longer-term evolutionary trends.

\section{Summary and conclusions} \label{Summary}

We presented a detailed case study of the well-known, strongly lensed Ly$\alpha$-emitting galaxy HCM 6A at $z = 6.5676$, located behind Abell 370 and observed as part of the {\it JWST}/CANUCS survey. Combining 19-band {\it HST}+NIRCam photometry, three {\it JWST}/NIRSpec slits, and NIRISS slitless spectroscopy, we constrained the galaxy’s Ly$\alpha$ emission, spatially resolved physical properties, and dust attenuation curve. A high-quality Ly$\alpha$ map from \texttt{SLEUTH} further revealed the spatial distribution of Ly$\alpha$ emission. Using a customized \texttt{BAGPIPES} SED-fitting framework with a flexible dust law, we derived stellar, nebular, and dust properties across multiple spatial scales—from integrated ($\approx$ kpc) to slit ($\approx$0.1 kpc) and pixel levels (down to $\approx$25 pc tangentially) - enabled by strong lensing magnification ($\mu \approx 8.3$–9.1). 

Together, these multi-wavelength, spatially resolved observations provide an unprecedented view of the interplay between dust, Ly$\alpha$ escape, and feedback in a clumpy galaxy near the end of reionization.
Main results are:
\begin{itemize}
    \item From integrated {\it HST}+NIRCam photometry, we infer an unlensed stellar mass of $\log M_\ast = 8.3$–$8.4$, consistent with the sum of the delensed masses of the three NIRSpec slit regions. Net, using the F115W continuum, which traces the rest-frame UV at 1500 Å, we measure an intrinsic UV magnitude of $M_{\rm UV} = -19.8 \pm 0.1$, corresponding to a luminosity of $L_\nu = 3.6\times10^{28}\ {\rm erg \ s^{-1}\ Hz^{-1}}$.
    \item Slit-level ($\approx0.1$ kpc) maps show that the more massive, older region S1 is moderately dusty and consistent across all attenuation indicators, suggesting a uniform ISM geometry. In contrast, the youngest region S3 shows strong discrepancies between stellar ($A_V$, $\beta$) and nebular ($A_V^{\rm B}$, line emission) tracers, indicating a complex, multiphase dust geometry, possibly shaped by recent feedback.
    \item Pixel-level, NIRISS Ly$\alpha$ emission line map shows that Ly$\alpha$ emission arises mainly from clump C3, where $A_V$ is also high on average. Pixel-level maps reveal that Ly$\alpha$ emerges from the dust-cleared core, while $A_V$ peaks in the clump outskirts. This is consistent with a possible recent ($\langle a \rangle_*^{\rm m} \lesssim 10$ Myr) starburst that ionized the core and expelled dust via radiation-driven outflows, enabling Ly$\alpha$ escape through low-opacity channels.
    \item We construct one of the first spatially resolved attenuation curve maps at $z > 6$. Slit-level measurements yield a broadly Calzetti-like slope ($S \approx 2.3$–2.6) and a UV bump strength ($B$) that increases with stellar age and lower $A_V$. Pixel-level maps show that both $S$ and $B$ rise in the more evolved region between clumps C1–C2, consistent with dust-grain processing and the formation of small (carbon-rich) grains in a star-forming region.
    \item We find weak ($\sim 2.6\sigma$) evidence for a UV bump in S1 ($\sim$25\% of the Milky Way bump).
\end{itemize}

These findings provide the spatially resolved view of HCM 6A down to $\approx 25$ pc, revealing how dust, gas, and stars interact within a moderately dusty LAE at the EoR. Our combination of slit-level photometry + spectroscopy and pixel-level photometry uncovers a complex, multiphase ISM shaped by recent star formation and feedback, and demonstrates that low-opacity channels enable Ly$\alpha$ escape.
Importantly, these results show that dust does not necessarily suppress Ly$\alpha$ escape; instead, the small-scale geometry of dust and stars—likely reshaped by recent feedback—plays a central role in regulating Ly$\alpha$ transmission.

However, some key physical quantities—such as $A_V^{\rm B}$, gas-phase metallicity, and ionization state—remain limited by the lack of fully resolved spectroscopy, while parameters like stellar age, attenuation, and $\beta$ remain uncertain due to low S/N at the pixel level. Future deep JWST/NIRSpec IFU observations can overcome these limitations by providing spatially resolved spectroscopy, which enables robust measurements of the dust, gas, and ionization structure, and allows for a definitive test of the mechanisms regulating Ly$\alpha$ escape.

\begin{acknowledgements}    
       We thank the anonymous referee for their valuable comments and suggestions, which helped improve the clarity and quality of this work.
       We thank Johannes Zabl for detailed comments and helpful suggestions during the preparation of this manuscript.
       VM, MB, GR, JJ, and NM acknowledge support from the ERC Grant FIRSTLIGHT and the Slovenian National Research Agency ARRS through grants N1-0238, P1-0188.
        GR and MB acknowledge support from the European Space Agency through Prodex Experiment Arrangement No. 4000146646.
        This research was enabled by grants 18JWST-GTO1 and 23JWGO2B04 from the Canadian Space Agency. 
        MS, VEC, GD, and GN acknowledge support from the Natural Sciences and Engineering Research Council (NSERC) of Canada through grants RGPIN-2020-06023, RGPAS-2020-00065.
       GN acknowledges support by the Canadian Space Agency under a contract with NRC Herzberg Astronomy and Astrophysics.
       AH acknowledges support by the Science and Technology Facilities Council (STFC), by the ERC through Advanced Grant 695671 “QUENCH”, and by the UKRI Frontier Research grant RISEandFALL.
       The authors acknowledge the use of the Canadian Advanced Network for Astronomy Research (CANFAR) Science Platform operated by the Canadian Astronomy Data Centre (CADC) and the Digital Research Alliance of Canada (DRAC), with support from the National Research Council of Canada s(NRC), the Canadian Space Agency (CSA), CANARIE, and the Canadian Foundation for Innovation (CFI).
       This work made use of Astropy:\footnote{\url{http://www.astropy.org}}, a community-developed core Python package and an ecosystem of tools and resources for astronomy \citep{2022ApJ...935..167A}.
\end{acknowledgements}

\bibliographystyle{aa} 
\bibliography{biblio} 

@software{Brammer_msaexp_NIRSpec_analyis_2022,
author = {Brammer, Gabriel},
doi = {10.5281/zenodo.7299500},
month = nov,
title = {{msaexp: NIRSpec analyis tools}},
url = {https://github.com/gbrammer/msaexp},
version = {0.3},
year = {2022}
}

@ARTICLE{2023MNRAS.522.3986F,
       author = {{Ferrara}, Andrea and {Pallottini}, Andrea and {Dayal}, Pratika},
        title = "{On the stunning abundance of super-early, luminous galaxies revealed by JWST}",
      journal = {\mnras},
         year = 2023,
        month = jul,
       volume = {522},
       number = {3},
        pages = {3986-3991},
          doi = {10.1093/mnras/stad1095},
archivePrefix = {arXiv},
       eprint = {2208.00720},
 primaryClass = {astro-ph.GA},
       adsurl = {https://ui.adsabs.harvard.edu/abs/2023MNRAS.522.3986F}
}

@ARTICLE{2024arXiv241114532S,
       author = {{Saxena}, Aayush and {Cameron}, Alex J. and {Katz}, Harley and {Bunker}, Andrew J. and {Chevallard}, Jacopo and {D'Eugenio}, Francesco and {Arribas}, Santiago and {Bhatawdekar}, Rachana and {Boyett}, Kristan and {Cargile}, Phillip A. and {Carniani}, Stefano and {Charlot}, Stephane and {Curti}, Mirko and {Curtis-Lake}, Emma and {Hainline}, Kevin and {Ji}, Zhiyuan and {Johnson}, Benjamin D. and {Jones}, Gareth C. and {Kumari}, Nimisha and {Laseter}, Isaac and {Maseda}, Michael V. and {Robertson}, Brant and {Simmonds}, Charlotte and {Tacchella}, Sandro and {Ubler}, Hannah and {Williams}, Christina C. and {Willott}, Chris and {Witstok}, Joris and {Zhu}, Yongda},
        title = "{Hitting the slopes: A spectroscopic view of UV continuum slopes of galaxies reveals a reddening at z > 9.5}",
      journal = {arXiv e-prints},
         year = 2024,
        month = nov,
          eid = {arXiv:2411.14532},
        pages = {arXiv:2411.14532},
          doi = {10.48550/arXiv.2411.14532},
archivePrefix = {arXiv},
       note = {submitted to MNRAS},
       eprint = {2411.14532},
 primaryClass = {astro-ph.GA},
       adsurl = {https://ui.adsabs.harvard.edu/abs/2024arXiv241114532S}
}

@ARTICLE{2011Ap&SS.331....1W,
       author = {{Walcher}, Jakob and {Groves}, Brent and {Budav{\'a}ri}, Tam{\'a}s and {Dale}, Daniel},
        title = "{Fitting the integrated spectral energy distributions of galaxies}",
      journal = {\apss},
         year = 2011,
        month = jan,
       volume = {331},
       number = {1},
        pages = {1-51},
          doi = {10.1007/s10509-010-0458-z},
archivePrefix = {arXiv},
       eprint = {1008.0395},
 primaryClass = {astro-ph.CO},
       adsurl = {https://ui.adsabs.harvard.edu/abs/2011Ap&SS.331....1W}
}

@ARTICLE{2024A&A...681A..94N,
       author = {{Nersesian}, Angelos and {van der Wel}, Arjen and {Gallazzi}, Anna and {Leja}, Joel and {Bezanson}, Rachel and {Bell}, Eric F. and {D'Eugenio}, Francesco and {de Graaff}, Anna and {Kaushal}, Yasha and {Martorano}, Marco and {Maseda}, Michael and {Zibetti}, Stefano},
        title = "{Less is less: Photometry alone cannot predict the observed spectral indices of z   1 galaxies from the LEGA-C spectroscopic survey}",
      journal = {\aap},
         year = 2024,
        month = jan,
       volume = {681},
          eid = {A94},
        pages = {A94},
          doi = {10.1051/0004-6361/202346769},
archivePrefix = {arXiv},
       eprint = {2310.18000},
 primaryClass = {astro-ph.GA},
       adsurl = {https://ui.adsabs.harvard.edu/abs/2024A&A...681A..94N}
}

@ARTICLE{2020MNRAS.495.2088C,
       author = {{Chaves-Montero}, Jon{\'a}s and {Hearin}, Andrew},
        title = "{Surrogate modelling the Baryonic Universe - I. The colour of star formation}",
      journal = {\mnras},
         year = 2020,
        month = jun,
       volume = {495},
       number = {2},
        pages = {2088-2104},
          doi = {10.1093/mnras/staa1230},
archivePrefix = {arXiv},
       eprint = {1910.11883},
 primaryClass = {astro-ph.GA},
       adsurl = {https://ui.adsabs.harvard.edu/abs/2020MNRAS.495.2088C}
}

@ARTICLE{2025A&A...695A..86N,
       author = {{Nersesian}, Angelos and {van der Wel}, Arjen and {Gallazzi}, Anna R. and {Kaushal}, Yasha and {Bezanson}, Rachel and {Zibetti}, Stefano and {Bell}, Eric F. and {D'Eugenio}, Francesco and {Leja}, Joel and {Martorano}, Marco and {Wu}, Po-Feng},
        title = "{More is better: Strong constraints on the stellar properties of LEGA-C z {\ensuremath{\sim}} 1 galaxies with Prospector}",
      journal = {\aap},
         year = 2025,
        month = mar,
       volume = {695},
          eid = {A86},
        pages = {A86},
          doi = {10.1051/0004-6361/202452662},
archivePrefix = {arXiv},
       eprint = {2502.03021},
 primaryClass = {astro-ph.GA},
       adsurl = {https://ui.adsabs.harvard.edu/abs/2025A&A...695A..86N}
}

@ARTICLE{2023MNRAS.526.4801T,
       author = {{Tsuna}, Daichi and {Nakazato}, Yurina and {Hartwig}, Tilman},
        title = "{A photon burst clears the earliest dusty galaxies: modelling dust in high-redshift galaxies from ALMA to JWST}",
      journal = {\mnras},
         year = 2023,
        month = dec,
       volume = {526},
       number = {4},
        pages = {4801-4813},
          doi = {10.1093/mnras/stad3043},
archivePrefix = {arXiv},
       eprint = {2309.02415},
 primaryClass = {astro-ph.GA},
       adsurl = {https://ui.adsabs.harvard.edu/abs/2023MNRAS.526.4801T}
}

@ARTICLE{2009A&A...507.1327H,
       author = {{Hunt}, L.~K. and {Hirashita}, H.},
        title = "{The size-density relation of extragalactic H II regions}",
      journal = {\aap},
         year = 2009,
        month = dec,
       volume = {507},
       number = {3},
        pages = {1327-1343},
          doi = {10.1051/0004-6361/200912020},
archivePrefix = {arXiv},
       eprint = {0910.2804},
 primaryClass = {astro-ph.CO},
       adsurl = {https://ui.adsabs.harvard.edu/abs/2009A&A...507.1327H}
}

@ARTICLE{2025NatAs...9..458M,
       author = {{Markov}, Vladan and {Gallerani}, Simona and {Ferrara}, Andrea and {Pallottini}, Andrea and {Parlanti}, Eleonora and {Mascia}, Fabio Di and {Sommovigo}, Laura and {Kohandel}, Mahsa},
        title = "{The evolution of dust attenuation in z {\ensuremath{\approx}} 2-12 galaxies observed by JWST}",
      journal = {Nature Astronomy},
         year = 2025,
        month = mar,
       volume = {9},
        pages = {458-468},
          doi = {10.1038/s41550-024-02426-1},
archivePrefix = {arXiv},
       eprint = {2402.05996},
 primaryClass = {astro-ph.GA},
       adsurl = {https://ui.adsabs.harvard.edu/abs/2025NatAs...9..458M}
}

@ARTICLE{2023ApJ...954..157S,
       author = {{Shapley}, Alice E. and {Sanders}, Ryan L. and {Reddy}, Naveen A. and {Topping}, Michael W. and {Brammer}, Gabriel B.},
        title = "{JWST/NIRSpec Balmer-line Measurements of Star Formation and Dust Attenuation at z   3-6}",
      journal = {\apj},
         year = 2023,
        month = sep,
       volume = {954},
       number = {2},
          eid = {157},
        pages = {157},
          doi = {10.3847/1538-4357/acea5a},
archivePrefix = {arXiv},
       eprint = {2301.03241},
 primaryClass = {astro-ph.GA},
       adsurl = {https://ui.adsabs.harvard.edu/abs/2023ApJ...954..157S}
}

@ARTICLE{2023ApJ...949L..11M,
       author = {{Matharu}, Jasleen and {Muzzin}, Adam and {Sarrouh}, Ghassan T.~E. and {Brammer}, Gabriel and {Abraham}, Roberto and {Asada}, Yoshihisa and {Brada{\v{c}}}, Maru{\v{s}}a and {Desprez}, Guillaume and {Martis}, Nicholas and {Mowla}, Lamiya and {Noirot}, Ga{\"e}l and {Sawicki}, Marcin and {Strait}, Victoria and {Willott}, Chris J. and {Gould}, Katriona M.~L. and {Grindlay}, Tess and {Harshan}, Anishya T.},
        title = "{A First Look at Spatially Resolved Balmer Decrements at 1.0 < z < 2.4 from JWST NIRISS Slitless Spectroscopy}",
      journal = {\apjl},
         year = 2023,
        month = may,
       volume = {949},
       number = {1},
          eid = {L11},
        pages = {L11},
          doi = {10.3847/2041-8213/acd1db},
archivePrefix = {arXiv},
       eprint = {2303.17624},
 primaryClass = {astro-ph.GA},
       adsurl = {https://ui.adsabs.harvard.edu/abs/2023ApJ...949L..11M}
}

@ARTICLE{2019OJAp....2E..10F,
       author = {{Feroz}, Farhan and {Hobson}, Michael P. and {Cameron}, Ewan and {Pettitt}, Anthony N.},
        title = "{Importance Nested Sampling and the MultiNest Algorithm}",
      journal = {The Open Journal of Astrophysics},
         year = 2019,
        month = nov,
       volume = {2},
       number = {1},
          eid = {10},
        pages = {10},
          doi = {10.21105/astro.1306.2144},
archivePrefix = {arXiv},
       eprint = {1306.2144},
 primaryClass = {astro-ph.IM},
       adsurl = {https://ui.adsabs.harvard.edu/abs/2019OJAp....2E..10F}
}

@article{Willott_2022,
doi = {10.1088/1538-3873/ac5158},
url = {https://dx.doi.org/10.1088/1538-3873/ac5158},
year = {2022},
month = {feb},
publisher = {The Astronomical Society of the Pacific},
volume = {134},
number = {1032},
pages = {025002},
author = {Willott, Chris J. and Doyon, René and Albert, Loic and Brammer, Gabriel B. and Dixon, William V. and Muzic, Koraljka and Ravindranath, Swara and Scholz, Aleks and Abraham, Roberto and Artigau, Étienne and Bradač, Maruša and Goudfrooij, Paul and Hutchings, John B. and Iyer, Kartheik G. and Jayawardhana, Ray and LaMassa, Stephanie and Martis, Nicholas and Meyer, Michael R. and Morishita, Takahiro and Mowla, Lamiya and Muzzin, Adam and Noirot, Gaël and Pacifici, Camilla and Rowlands, Neil and Sarrouh, Ghassan and Sawicki, Marcin and Taylor, Joanna M. and Volk, Kevin and Zabl, Johannes},
title = {The Near-infrared Imager and Slitless Spectrograph for the James Webb Space Telescope. II. Wide Field Slitless Spectroscopy},
journal = {Publications of the Astronomical Society of the Pacific}
}

@ARTICLE{2011ApJ...730....8H,
       author = {{Hayes}, Matthew and {Schaerer}, Daniel and {{\"O}stlin}, G{\"o}ran and {Mas-Hesse}, J. Miguel and {Atek}, Hakim and {Kunth}, Daniel},
        title = "{On the Redshift Evolution of the Ly{\ensuremath{\alpha}} Escape Fraction and the Dust Content of Galaxies}",
      journal = {\apj},
         year = 2011,
        month = mar,
       volume = {730},
       number = {1},
          eid = {8},
        pages = {8},
          doi = {10.1088/0004-637X/730/1/8},
archivePrefix = {arXiv},
       eprint = {1010.4796},
 primaryClass = {astro-ph.CO},
       adsurl = {https://ui.adsabs.harvard.edu/abs/2011ApJ...730....8H}
}

@ARTICLE{2020ARA&A..58..617O,
       author = {{Ouchi}, Masami and {Ono}, Yoshiaki and {Shibuya}, Takatoshi},
        title = "{Observations of the Lyman-{\ensuremath{\alpha}} Universe}",
      journal = {\araa},
         year = 2020,
        month = aug,
       volume = {58},
        pages = {617-659},
          doi = {10.1146/annurev-astro-032620-021859},
archivePrefix = {arXiv},
       eprint = {2012.07960},
 primaryClass = {astro-ph.GA},
       adsurl = {https://ui.adsabs.harvard.edu/abs/2020ARA&A..58..617O}
}

@ARTICLE{2014ApJ...794....5T,
       author = {{Tilvi}, V. and {Papovich}, C. and {Finkelstein}, S.~L. and {Long}, J. and {Song}, M. and {Dickinson}, M. and {Ferguson}, H.~C. and {Koekemoer}, A.~M. and {Giavalisco}, M. and {Mobasher}, B.},
        title = "{Rapid Decline of Ly{\ensuremath{\alpha}} Emission toward the Reionization Era}",
      journal = {\apj},
         year = 2014,
        month = oct,
       volume = {794},
       number = {1},
          eid = {5},
        pages = {5},
          doi = {10.1088/0004-637X/794/1/5},
archivePrefix = {arXiv},
       eprint = {1405.4869},
 primaryClass = {astro-ph.CO},
       adsurl = {https://ui.adsabs.harvard.edu/abs/2014ApJ...794....5T}
}

@ARTICLE{2013ApJ...775L..29T,
       author = {{Treu}, Tommaso and {Schmidt}, Kasper B. and {Trenti}, Michele and {Bradley}, Larry D. and {Stiavelli}, Massimo},
        title = "{The Changing Ly{\ensuremath{\alpha}} Optical Depth in the Range 6 < z < 9 from the MOSFIRE Spectroscopy of Y-dropouts}",
      journal = {\apjl},
         year = 2013,
        month = sep,
       volume = {775},
       number = {1},
          eid = {L29},
        pages = {L29},
          doi = {10.1088/2041-8205/775/1/L29},
archivePrefix = {arXiv},
       eprint = {1308.5985},
 primaryClass = {astro-ph.CO},
       adsurl = {https://ui.adsabs.harvard.edu/abs/2013ApJ...775L..29T}
}

@ARTICLE{2024MNRAS.531L..34K,
       author = {{Keating}, Laura C. and {Puchwein}, Ewald and {Bolton}, James S. and {Haehnelt}, Martin G. and {Kulkarni}, Girish},
        title = "{The origin of the characteristic shape and scatter of intergalactic damping wings during reionization}",
      journal = {\mnras},
         year = 2024,
        month = jun,
       volume = {531},
       number = {1},
        pages = {L34-L39},
          doi = {10.1093/mnrasl/slae022},
archivePrefix = {arXiv},
       eprint = {2308.11709},
 primaryClass = {astro-ph.CO},
       adsurl = {https://ui.adsabs.harvard.edu/abs/2024MNRAS.531L..34K}
}

@ARTICLE{2024MNRAS.532.1646K,
       author = {{Keating}, Laura C. and {Bolton}, James S. and {Cullen}, Fergus and {Haehnelt}, Martin G. and {Puchwein}, Ewald and {Kulkarni}, Girish},
        title = "{JWST observations of galaxy-damping wings during reionization interpreted with cosmological simulations}",
      journal = {\mnras},
         year = 2024,
        month = aug,
       volume = {532},
       number = {2},
        pages = {1646-1658},
          doi = {10.1093/mnras/stae1530},
archivePrefix = {arXiv},
       eprint = {2308.05800},
 primaryClass = {astro-ph.GA},
       adsurl = {https://ui.adsabs.harvard.edu/abs/2024MNRAS.532.1646K}
}

@ARTICLE{2024A&A...689A..44T,
       author = {{Torralba}, Alberto and {Matthee}, Jorryt and {Naidu}, Rohan P. and {Mackenzie}, Ruari and {Pezzulli}, Gabriele and {Hutter}, Anne and {Arnalte-Mur}, Pablo and {Gurung-L{\'o}pez}, Siddhartha and {Tacchella}, Sandro and {Oesch}, Pascal and {Kashino}, Daichi and {Conroy}, Charlie and {Sobral}, David},
        title = "{Anatomy of an ionized bubble: NIRCam grism spectroscopy of the z = 6.6 double-peaked Lyman-{\ensuremath{\alpha}} emitter COLA1 and its environment}",
      journal = {\aap},
         year = 2024,
        month = sep,
       volume = {689},
          eid = {A44},
        pages = {A44},
          doi = {10.1051/0004-6361/202450318},
archivePrefix = {arXiv},
       eprint = {2404.10040},
 primaryClass = {astro-ph.GA},
       adsurl = {https://ui.adsabs.harvard.edu/abs/2024A&A...689A..44T}
}

@ARTICLE{2022MNRAS.515.5790L,
       author = {{Leonova}, E. and {Oesch}, P.~A. and {Qin}, Y. and {Naidu}, R.~P. and {Wyithe}, J.~S.~B. and {de Barros}, S. and {Bouwens}, R.~J. and {Ellis}, R.~S. and {Endsley}, R.~M. and {Hutter}, A. and {Illingworth}, G.~D. and {Kerutt}, J. and {Labb{\'e}}, I. and {Laporte}, N. and {Magee}, D. and {Mutch}, S.~J. and {Roberts-Borsani}, G.~W. and {Smit}, R. and {Stark}, D.~P. and {Stefanon}, M. and {Tacchella}, S. and {Zitrin}, A.},
        title = "{The prevalence of galaxy overdensities around UV-luminous Lyman �� emitters in the Epoch of Reionization}",
      journal = {\mnras},
         year = 2022,
        month = oct,
       volume = {515},
       number = {4},
        pages = {5790-5801},
          doi = {10.1093/mnras/stac1908},
archivePrefix = {arXiv},
       eprint = {2112.07675},
 primaryClass = {astro-ph.GA},
       adsurl = {https://ui.adsabs.harvard.edu/abs/2022MNRAS.515.5790L}
}

@ARTICLE{2024A&A...682A..40W,
       author = {{Witstok}, Joris and {Smit}, Renske and {Saxena}, Aayush and {Jones}, Gareth C. and {Helton}, Jakob M. and {Sun}, Fengwu and {Maiolino}, Roberto and {Kumari}, Nimisha and {Stark}, Daniel P. and {Bunker}, Andrew J. and {Arribas}, Santiago and {Baker}, William M. and {Bhatawdekar}, Rachana and {Boyett}, Kristan and {Cameron}, Alex J. and {Carniani}, Stefano and {Charlot}, Stephane and {Chevallard}, Jacopo and {Curti}, Mirko and {Curtis-Lake}, Emma and {Eisenstein}, Daniel J. and {Endsley}, Ryan and {Hainline}, Kevin and {Ji}, Zhiyuan and {Johnson}, Benjamin D. and {Looser}, Tobias J. and {Nelson}, Erica and {Perna}, Michele and {Rix}, Hans-Walter and {Robertson}, Brant E. and {Sandles}, Lester and {Scholtz}, Jan and {Simmonds}, Charlotte and {Tacchella}, Sandro and {{\"U}bler}, Hannah and {Williams}, Christina C. and {Willmer}, Christopher N.~A. and {Willott}, Chris},
        title = "{Inside the bubble: exploring the environments of reionisation-era Lyman-{\ensuremath{\alpha}} emitting galaxies with JADES and FRESCO}",
      journal = {\aap},
         year = 2024,
        month = feb,
       volume = {682},
          eid = {A40},
        pages = {A40},
          doi = {10.1051/0004-6361/202347176},
archivePrefix = {arXiv},
       eprint = {2306.04627},
 primaryClass = {astro-ph.GA},
       adsurl = {https://ui.adsabs.harvard.edu/abs/2024A&A...682A..40W}
}

@ARTICLE{2018ApJ...856....2M,
       author = {{Mason}, Charlotte A. and {Treu}, Tommaso and {Dijkstra}, Mark and {Mesinger}, Andrei and {Trenti}, Michele and {Pentericci}, Laura and {de Barros}, Stephane and {Vanzella}, Eros},
        title = "{The Universe Is Reionizing at z {\ensuremath{\sim}} 7: Bayesian Inference of the IGM Neutral Fraction Using Ly{\ensuremath{\alpha}} Emission from Galaxies}",
      journal = {\apj},
         year = 2018,
        month = mar,
       volume = {856},
       number = {1},
          eid = {2},
        pages = {2},
          doi = {10.3847/1538-4357/aab0a7},
archivePrefix = {arXiv},
       eprint = {1709.05356},
 primaryClass = {astro-ph.CO},
       adsurl = {https://ui.adsabs.harvard.edu/abs/2018ApJ...856....2M}
}

@ARTICLE{2004MNRAS.349.1137S,
       author = {{Santos}, Michael R.},
        title = "{Probing reionization with Lyman {\ensuremath{\alpha}} emission lines}",
      journal = {\mnras},
         year = 2004,
        month = apr,
       volume = {349},
       number = {3},
        pages = {1137-1152},
          doi = {10.1111/j.1365-2966.2004.07594.x},
archivePrefix = {arXiv},
       eprint = {astro-ph/0308196},
 primaryClass = {astro-ph},
       adsurl = {https://ui.adsabs.harvard.edu/abs/2004MNRAS.349.1137S}
}

@ARTICLE{1991ApJ...370L..85N,
       author = {{Neufeld}, David A.},
        title = "{The Escape of Lyman-Alpha Radiation from a Multiphase Interstellar Medium}",
      journal = {\apjl},
         year = 1991,
        month = apr,
       volume = {370},
        pages = {L85},
          doi = {10.1086/185983},
       adsurl = {https://ui.adsabs.harvard.edu/abs/1991ApJ...370L..85N}
}

@ARTICLE{2002ApJ...568L..75H,
       author = {{Hu}, E.~M. and {Cowie}, L.~L. and {McMahon}, R.~G. and {Capak}, P. and {Iwamuro}, F. and {Kneib}, J. -P. and {Maihara}, T. and {Motohara}, K.},
        title = "{A Redshift z=6.56 Galaxy behind the Cluster Abell 370}",
      journal = {\apjl},
         year = 2002,
        month = apr,
       volume = {568},
       number = {2},
        pages = {L75-L79},
          doi = {10.1086/340424},
archivePrefix = {arXiv},
       eprint = {astro-ph/0203091},
 primaryClass = {astro-ph},
       adsurl = {https://ui.adsabs.harvard.edu/abs/2002ApJ...568L..75H}
}

@ARTICLE{2007A&A...475..513B,
       author = {{Boone}, F. and {Schaerer}, D. and {Pell{\'o}}, R. and {Combes}, F. and {Egami}, E.},
        title = "{Millimeter observations of HCM 6A, a gravitationally lensed Ly{\ensuremath{\alpha}} emitting galaxy at z = 6.56}",
      journal = {\aap},
         year = 2007,
        month = nov,
       volume = {475},
       number = {2},
        pages = {513-517},
          doi = {10.1051/0004-6361:20078253},
archivePrefix = {arXiv},
       eprint = {0709.3721},
 primaryClass = {astro-ph},
       adsurl = {https://ui.adsabs.harvard.edu/abs/2007A&A...475..513B}
}

@ARTICLE{2020ApJ...896..156F,
       author = {{Fuller}, S. and {Lemaux}, B.~C. and {Brada{\v{c}}}, M. and {Hoag}, A. and {Schmidt}, K.~B. and {Huang}, K. and {Strait}, V. and {Mason}, C. and {Treu}, T. and {Pentericci}, L. and {Trenti}, M. and {Henry}, A. and {Malkan}, M.},
        title = "{Spectroscopically Confirmed Ly{\ensuremath{\alpha}} Emitters from Redshift 5 to 7 behind 10 Galaxy Cluster Lenses}",
      journal = {\apj},
         year = 2020,
        month = jun,
       volume = {896},
       number = {2},
          eid = {156},
        pages = {156},
          doi = {10.3847/1538-4357/ab959f},
archivePrefix = {arXiv},
       eprint = {2002.08952},
 primaryClass = {astro-ph.CO},
       adsurl = {https://ui.adsabs.harvard.edu/abs/2020ApJ...896..156F}
}

@ARTICLE{2000ApJ...543L.119B,
       author = {{Bautz}, M.~W. and {Malm}, M.~R. and {Baganoff}, F.~K. and {Ricker}, G.~R. and {Canizares}, C.~R. and {Brandt}, W.~N. and {Hornschemeier}, A.~E. and {Garmire}, G.~P.},
        title = "{Detection of X-Ray Emission from Gravitationally Lensed Submillimeter Sources in the Field of Abell 370}",
      journal = {\apjl},
         year = 2000,
        month = nov,
       volume = {543},
       number = {2},
        pages = {L119-L123},
          doi = {10.1086/317272},
archivePrefix = {arXiv},
       eprint = {astro-ph/0008050},
 primaryClass = {astro-ph},
       adsurl = {https://ui.adsabs.harvard.edu/abs/2000ApJ...543L.119B}
}

@ARTICLE{2014PhR...541...45C,
       author = {{Casey}, Caitlin M. and {Narayanan}, Desika and {Cooray}, Asantha},
        title = "{Dusty star-forming galaxies at high redshift}",
      journal = {\physrep},
         year = 2014,
        month = aug,
       volume = {541},
       number = {2},
        pages = {45-161},
          doi = {10.1016/j.physrep.2014.02.009},
archivePrefix = {arXiv},
       eprint = {1402.1456},
 primaryClass = {astro-ph.CO},
       adsurl = {https://ui.adsabs.harvard.edu/abs/2014PhR...541...45C}
}

@ARTICLE{2000MNRAS.315..115D,
       author = {{Dunne}, Loretta and {Eales}, Stephen and {Edmunds}, Michael and {Ivison}, Rob and {Alexander}, Paul and {Clements}, David L.},
        title = "{The SCUBA Local Universe Galaxy Survey - I. First measurements of the submillimetre luminosity and dust mass functions}",
      journal = {\mnras},
         year = 2000,
        month = jun,
       volume = {315},
       number = {1},
        pages = {115-139},
          doi = {10.1046/j.1365-8711.2000.03386.x},
archivePrefix = {arXiv},
       eprint = {astro-ph/0002234},
 primaryClass = {astro-ph},
       adsurl = {https://ui.adsabs.harvard.edu/abs/2000MNRAS.315..115D}
}

@ARTICLE{2012MNRAS.425.3094C,
       author = {{Casey}, Caitlin M.},
        title = "{Far-infrared spectral energy distribution fitting for galaxies near and far}",
      journal = {\mnras},
         year = 2012,
        month = oct,
       volume = {425},
       number = {4},
        pages = {3094-3103},
          doi = {10.1111/j.1365-2966.2012.21455.x},
archivePrefix = {arXiv},
       eprint = {1206.1595},
 primaryClass = {astro-ph.CO},
       adsurl = {https://ui.adsabs.harvard.edu/abs/2012MNRAS.425.3094C}
}

@ARTICLE{2024ApJS..275...36F,
       author = {{Fujimoto}, Seiji and {Kohno}, Kotaro and {Ouchi}, Masami and {Oguri}, Masamune and {Kokorev}, Vasily and {Brammer}, Gabriel and {Sun}, Fengwu and {Gonz{\'a}lez-L{\'o}pez}, Jorge and {Bauer}, Franz E. and {Caminha}, Gabriel B. and {Hatsukade}, Bunyo and {Richard}, Johan and {Smail}, Ian and {Tsujita}, Akiyoshi and {Ueda}, Yoshihiro and {Uematsu}, Ryosuke and {Zitrin}, Adi and {Coe}, Dan and {Kneib}, Jean-Paul and {Postman}, Marc and {Umetsu}, Keiichi and {Lagos}, Claudia del P. and {Popping}, Gerg{\"o} and {Ao}, Yiping and {Bradley}, Larry and {Caputi}, Karina and {Dessauges-Zavadsky}, Miroslava and {Egami}, Eiichi and {Espada}, Daniel and {Ivison}, R.~J. and {Jauzac}, Mathilde and {Knudsen}, Kirsten K. and {Koekemoer}, Anton M. and {Magdis}, Georgios E. and {Mahler}, Guillaume and {Mu{\~n}oz Arancibia}, A.~M. and {Rawle}, Timothy and {Shimasaku}, Kazuhiro and {Toft}, Sune and {Umehata}, Hideki and {Valentino}, Francesco and {Wang}, Tao and {Wang}, Wei-Hao},
        title = "{ALMA Lensing Cluster Survey: Deep 1.2 mm Number Counts and Infrared Luminosity Functions at z ≃ 1{\textendash}8}",
      journal = {\apjs},
         year = 2024,
        month = dec,
       volume = {275},
       number = {2},
          eid = {36},
        pages = {36},
          doi = {10.3847/1538-4365/ad5ae2},
archivePrefix = {arXiv},
       eprint = {2303.01658},
 primaryClass = {astro-ph.GA},
       adsurl = {https://ui.adsabs.harvard.edu/abs/2024ApJS..275...36F}
}

@ARTICLE{2005ApJ...635L...5C,
       author = {{Chary}, Ranga-Ram and {Stern}, Daniel and {Eisenhardt}, Peter},
        title = "{Spitzer Constraints on the z = 6.56 Galaxy Lensed by Abell 370}",
      journal = {\apjl},
         year = 2005,
        month = dec,
       volume = {635},
       number = {1},
        pages = {L5-L8},
          doi = {10.1086/499205},
archivePrefix = {arXiv},
       eprint = {astro-ph/0510827},
 primaryClass = {astro-ph},
       adsurl = {https://ui.adsabs.harvard.edu/abs/2005ApJ...635L...5C}
}

@ARTICLE{2013ApJ...771L..20K,
       author = {{Kanekar}, Nissim and {Wagg}, Jeff and {Chary}, Ranga Ram and {Carilli}, Christopher L.},
        title = "{A Search for C II 158 {\ensuremath{\mu}}m Line Emission in HCM 6A, a Ly{\ensuremath{\alpha}} Emitter at z = 6.56}",
      journal = {\apjl},
         year = 2013,
        month = jul,
       volume = {771},
       number = {2},
          eid = {L20},
        pages = {L20},
          doi = {10.1088/2041-8205/771/2/L20},
archivePrefix = {arXiv},
       eprint = {1305.6469},
 primaryClass = {astro-ph.CO},
       adsurl = {https://ui.adsabs.harvard.edu/abs/2013ApJ...771L..20K}
}

@ARTICLE{2020MNRAS.493.5120M,
       author = {{Maseda}, Michael V. and {Bacon}, Roland and {Lam}, Daniel and {Matthee}, Jorryt and {Brinchmann}, Jarle and {Schaye}, Joop and {Labbe}, Ivo and {Schmidt}, Kasper B. and {Boogaard}, Leindert and {Bouwens}, Rychard and {Cantalupo}, Sebastiano and {Franx}, Marijn and {Hashimoto}, Takuya and {Inami}, Hanae and {Kusakabe}, Haruka and {Mahler}, Guillaume and {Nanayakkara}, Themiya and {Richard}, Johan and {Wisotzki}, Lutz},
        title = "{Elevated ionizing photon production efficiency in faint high-equivalent-width Lyman-{\ensuremath{\alpha}} emitters}",
      journal = {\mnras},
         year = 2020,
        month = apr,
       volume = {493},
       number = {4},
        pages = {5120-5130},
          doi = {10.1093/mnras/staa622},
archivePrefix = {arXiv},
       eprint = {2002.11117},
 primaryClass = {astro-ph.GA},
       adsurl = {https://ui.adsabs.harvard.edu/abs/2020MNRAS.493.5120M}
}

@ARTICLE{2017ApJ...844..171Y,
       author = {{Yang}, Huan and {Malhotra}, Sangeeta and {Gronke}, Max and {Rhoads}, James E. and {Leitherer}, Claus and {Wofford}, Aida and {Jiang}, Tianxing and {Dijkstra}, Mark and {Tilvi}, V. and {Wang}, Junxian},
        title = "{Ly{\ensuremath{\alpha}} Profile, Dust, and Prediction of Ly{\ensuremath{\alpha}} Escape Fraction in Green Pea Galaxies}",
      journal = {\apj},
         year = 2017,
        month = aug,
       volume = {844},
       number = {2},
          eid = {171},
        pages = {171},
          doi = {10.3847/1538-4357/aa7d4d},
archivePrefix = {arXiv},
       eprint = {1701.01857},
 primaryClass = {astro-ph.GA},
       adsurl = {https://ui.adsabs.harvard.edu/abs/2017ApJ...844..171Y}
}

@ARTICLE{2009ApJ...704L..98S,
       author = {{Scarlata}, C. and {Colbert}, J. and {Teplitz}, H.~I. and {Panagia}, N. and {Hayes}, M. and {Siana}, B. and {Rau}, A. and {Francis}, P. and {Caon}, A. and {Pizzella}, A. and {Bridge}, C.},
        title = "{The Effect of Dust Geometry on the Ly{\ensuremath{\alpha}} Output of Galaxies}",
      journal = {\apjl},
         year = 2009,
        month = oct,
       volume = {704},
       number = {2},
        pages = {L98-L102},
          doi = {10.1088/0004-637X/704/2/L98},
archivePrefix = {arXiv},
       eprint = {0909.3847},
 primaryClass = {astro-ph.CO},
       adsurl = {https://ui.adsabs.harvard.edu/abs/2009ApJ...704L..98S}
}

@ARTICLE{2021MNRAS.505.1382M,
       author = {{Matthee}, Jorryt and {Sobral}, David and {Hayes}, Matthew and {Pezzulli}, Gabriele and {Gronke}, Max and {Schaerer}, Daniel and {Naidu}, Rohan P. and {R{\"o}ttgering}, Huub and {Calhau}, Jo{\~a}o and {Paulino-Afonso}, Ana and {Santos}, S{\'e}rgio and {Amor{\'\i}n}, Ricardo},
        title = "{The X-SHOOTER Lyman {\ensuremath{\alpha}} survey at z = 2 (XLS-z2) I: what makes a galaxy a Lyman {\ensuremath{\alpha}} emitter?}",
      journal = {\mnras},
         year = 2021,
        month = jul,
       volume = {505},
       number = {1},
        pages = {1382-1412},
          doi = {10.1093/mnras/stab1304},
archivePrefix = {arXiv},
       eprint = {2102.07779},
 primaryClass = {astro-ph.GA},
       adsurl = {https://ui.adsabs.harvard.edu/abs/2021MNRAS.505.1382M}
}

@ARTICLE{2016ApJ...826...14G,
       author = {{Gronke}, M. and {Dijkstra}, M.},
        title = "{Lyman-{\ensuremath{\alpha}} Spectra from Multiphase Outflows, and their Connection to Shell Models}",
      journal = {\apj},
         year = 2016,
        month = jul,
       volume = {826},
       number = {1},
          eid = {14},
        pages = {14},
          doi = {10.3847/0004-637X/826/1/14},
archivePrefix = {arXiv},
       eprint = {1604.06805},
 primaryClass = {astro-ph.GA},
       adsurl = {https://ui.adsabs.harvard.edu/abs/2016ApJ...826...14G}
}

@ARTICLE{2019MNRAS.486.2197B,
       author = {{Behrens}, C. and {Pallottini}, A. and {Ferrara}, A. and {Gallerani}, S. and {Vallini}, L.},
        title = "{Ly {\ensuremath{\alpha}} emission from galaxies in the Epoch of Reionization}",
      journal = {\mnras},
         year = 2019,
        month = jun,
       volume = {486},
       number = {2},
        pages = {2197-2209},
          doi = {10.1093/mnras/stz980},
archivePrefix = {arXiv},
       eprint = {1903.06185},
 primaryClass = {astro-ph.GA},
       adsurl = {https://ui.adsabs.harvard.edu/abs/2019MNRAS.486.2197B}
}

@ARTICLE{2010ApJ...711..693K,
       author = {{Kornei}, Katherine A. and {Shapley}, Alice E. and {Erb}, Dawn K. and {Steidel}, Charles C. and {Reddy}, Naveen A. and {Pettini}, Max and {Bogosavljevi{\'c}}, Milan},
        title = "{The Relationship between Stellar Populations and Ly{\ensuremath{\alpha}} Emission in Lyman Break Galaxies}",
      journal = {\apj},
         year = 2010,
        month = mar,
       volume = {711},
       number = {2},
        pages = {693-710},
          doi = {10.1088/0004-637X/711/2/693},
archivePrefix = {arXiv},
       eprint = {0911.2000},
 primaryClass = {astro-ph.CO},
       adsurl = {https://ui.adsabs.harvard.edu/abs/2010ApJ...711..693K}
}

@ARTICLE{2008A&A...491...89V,
       author = {{Verhamme}, A. and {Schaerer}, D. and {Atek}, H. and {Tapken}, C.},
        title = "{3D Ly{\ensuremath{\alpha}} radiation transfer. III. Constraints on gas and stellar properties of z \raisebox{-0.5ex}\textasciitilde 3 Lyman break galaxies (LBG) and implications for high-z LBGs and Ly{\ensuremath{\alpha}} emitters}",
      journal = {\aap},
         year = 2008,
        month = nov,
       volume = {491},
       number = {1},
        pages = {89-111},
          doi = {10.1051/0004-6361:200809648},
archivePrefix = {arXiv},
       eprint = {0805.3601},
 primaryClass = {astro-ph},
       adsurl = {https://ui.adsabs.harvard.edu/abs/2008A&A...491...89V}
}

@ARTICLE{2009ApJ...691..465F,
       author = {{Finkelstein}, Steven L. and {Rhoads}, James E. and {Malhotra}, Sangeeta and {Grogin}, Norman},
        title = "{Lyman Alpha Galaxies: Primitive, Dusty, or Evolved?}",
      journal = {\apj},
         year = 2009,
        month = jan,
       volume = {691},
       number = {1},
        pages = {465-481},
          doi = {10.1088/0004-637X/691/1/465},
archivePrefix = {arXiv},
       eprint = {0806.3269},
 primaryClass = {astro-ph},
       adsurl = {https://ui.adsabs.harvard.edu/abs/2009ApJ...691..465F}
}

@ARTICLE{2008ApJ...678..655F,
       author = {{Finkelstein}, Steven L. and {Rhoads}, James E. and {Malhotra}, Sangeeta and {Grogin}, Norman and {Wang}, Junxian},
        title = "{Effects of Dust Geometry in Ly{\ensuremath{\alpha}} Galaxies at z = 4.4}",
      journal = {\apj},
         year = 2008,
        month = may,
       volume = {678},
       number = {2},
        pages = {655-668},
          doi = {10.1086/525272},
archivePrefix = {arXiv},
       eprint = {0708.4226},
 primaryClass = {astro-ph},
       adsurl = {https://ui.adsabs.harvard.edu/abs/2008ApJ...678..655F}
}

@ARTICLE{2006MNRAS.367..979H,
       author = {{Hansen}, Matthew and {Oh}, S. Peng},
        title = "{Lyman {\ensuremath{\alpha}} radiative transfer in a multiphase medium}",
      journal = {\mnras},
         year = 2006,
        month = apr,
       volume = {367},
       number = {3},
        pages = {979-1002},
          doi = {10.1111/j.1365-2966.2005.09870.x},
archivePrefix = {arXiv},
       eprint = {astro-ph/0507586},
 primaryClass = {astro-ph},
       adsurl = {https://ui.adsabs.harvard.edu/abs/2006MNRAS.367..979H}
}

@ARTICLE{2009A&A...506L...1A,
       author = {{Atek}, H. and {Kunth}, D. and {Schaerer}, D. and {Hayes}, M. and {Deharveng}, J.~M. and {{\"O}stlin}, G. and {Mas-Hesse}, J.~M.},
        title = "{Empirical estimate of Ly{\ensuremath{\alpha}} escape fraction in a statistical sample of Ly{\ensuremath{\alpha}} emitters}",
      journal = {\aap},
         year = 2009,
        month = nov,
       volume = {506},
       number = {2},
        pages = {L1-L4},
          doi = {10.1051/0004-6361/200912787},
archivePrefix = {arXiv},
       eprint = {0906.5349},
 primaryClass = {astro-ph.CO},
       adsurl = {https://ui.adsabs.harvard.edu/abs/2009A&A...506L...1A}
}

@ARTICLE{2008A&A...488..491A,
       author = {{Atek}, H. and {Kunth}, D. and {Hayes}, M. and {{\"O}stlin}, G. and {Mas-Hesse}, J.~M.},
        title = "{On the detectability of Ly{\ensuremath{\alpha}} emission in star forming galaxies. The role of dust}",
      journal = {\aap},
         year = 2008,
        month = sep,
       volume = {488},
       number = {2},
        pages = {491-509},
          doi = {10.1051/0004-6361:200809527},
archivePrefix = {arXiv},
       eprint = {0805.3501},
 primaryClass = {astro-ph},
       adsurl = {https://ui.adsabs.harvard.edu/abs/2008A&A...488..491A}
}

@ARTICLE{2025A&A...694A.286F,
       author = {{Ferrara}, A. and {Pallottini}, A. and {Sommovigo}, L.},
        title = "{Blue monsters at z > 10: Where all their dust has gone}",
      journal = {\aap},
         year = 2025,
        month = feb,
       volume = {694},
          eid = {A286},
        pages = {A286},
          doi = {10.1051/0004-6361/202452707},
archivePrefix = {arXiv},
       eprint = {2410.19042},
 primaryClass = {astro-ph.GA},
       adsurl = {https://ui.adsabs.harvard.edu/abs/2025A&A...694A.286F}
}

@ARTICLE{2024A&A...688A.106N,
       author = {{Napolitano}, L. and {Pentericci}, L. and {Santini}, P. and {Calabr{\`o}}, A. and {Mascia}, S. and {Llerena}, M. and {Castellano}, M. and {Dickinson}, M. and {Finkelstein}, S.~L. and {Amor{\'\i}n}, R. and {Arrabal Haro}, P. and {Bagley}, M. and {Bhatawdekar}, R. and {Cleri}, N.~J. and {Davis}, K. and {Gardner}, J.~P. and {Gawiser}, E. and {Giavalisco}, M. and {Hathi}, N. and {Holwerda}, B.~W. and {Hu}, W. and {Jung}, I. and {Kartaltepe}, J.~S. and {Koekemoer}, A.~M. and {Larson}, R.~L. and {Merlin}, E. and {Mobasher}, B. and {Papovich}, C. and {Park}, H. and {Pirzkal}, N. and {Trump}, J.~R. and {Wilkins}, S.~M. and {Yung}, L.~Y.~A.},
        title = "{Peering into cosmic reionization: Ly{\ensuremath{\alpha}} visibility evolution from galaxies at z = 4.5-8.5 with JWST}",
      journal = {\aap},
         year = 2024,
        month = aug,
       volume = {688},
          eid = {A106},
        pages = {A106},
          doi = {10.1051/0004-6361/202449644},
archivePrefix = {arXiv},
       eprint = {2402.11220},
 primaryClass = {astro-ph.GA},
       adsurl = {https://ui.adsabs.harvard.edu/abs/2024A&A...688A.106N}
}

@misc{matharu_2022_7628094,
  author       = {Matharu, Jasleen and
                  Brammer, Gabriel},
  title        = {Updated Configuration files for JWST NIRISS
                   Slitless Spectroscopy
                  },
  month        = dec,
  year         = 2022,
  publisher    = {Zenodo},
  version      = {1.0},
  doi          = {10.5281/zenodo.7628094},
  url          = {https://doi.org/10.5281/zenodo.7628094},
}

@TECHREPORT{2025jwst.rept.9057N,
       author = {{Noirot}, Ga{\"e}l and {NIRISS WFSS Team} and {Goudfrooij}, Paul and {LaMassa}, Stephanie and {Plesha}, Rachel and {Taylor}, Jo and {Volk}, Kevin},
        title = "{Global sky background images for JWST/NIRISS Wide-Field Slitless Spectroscopy}",
  institution = {STScI},
         year = 2025,
       number = {Technical Report JWST-STScI-009057},
 howpublished = {Technical Report JWST-STScI-009057, 24 pages},
          doi = {10.48550/arXiv.2507.10650},
       adsurl = {https://ui.adsabs.harvard.edu/abs/2025jwst.rept.9057N}
}

@ARTICLE{2026arXiv260207347N,
       author = {{Nakazato}, Yurina and {Matsumoto}, Kosei and {Inoue}, Akio K. and {Ceverino}, Daniel and {Hosokawa}, Takashi and {Toyouchi}, Daisuke},
        title = "{Clump-Scale Dust Attenuation in Epoch of Reionization Galaxies: Spatially Resolved Properties from FirstLight Simulations}",
      journal = {arXiv e-prints},
         year = 2026,
        month = feb,
          eid = {arXiv:2602.07347},
        pages = {arXiv:2602.07347},
        doi = {10.48550/arXiv.2602.07347},
        archivePrefix = {arXiv},
        note  = {Submitted to ApJ},
       eprint = {2602.07347},
 primaryClass = {astro-ph.GA},
       adsurl = {https://ui.adsabs.harvard.edu/abs/2026arXiv260207347N}
}

@software{2021zndo...3377984C,
       author = {{Comrie}, Angus and {Wang}, Kuo-Song and {Hsu}, Shou-Chieh and {Moraghan}, Anthony and {Harris}, Pamela and {Pang}, Qi and {Pi{\'n}ska}, Adrianna and {Chiang}, Cheng-Chin and {Chang}, Tien-Hao and {Hwang}, Yu-Hsuan and {Jan}, Hengtai and {Lin}, Ming-Yi and {Simmonds}, Rob},
        title = "{CARTA: The Cube Analysis and Rendering Tool for Astronomy}",
         year = 2021,
        month = jun,
          eid = {10.5281/zenodo.3377984},
          doi = {10.5281/zenodo.3377984},
      version = {2.0.0},
    publisher = {Zenodo},
       adsurl = {https://ui.adsabs.harvard.edu/abs/2021zndo...3377984C}
}

@ARTICLE{2026ApJS..282....3S,
       author = {{Sarrouh}, Ghassan T.~E. and {Asada}, Yoshihisa and {Martis}, Nicholas S. and {Willott}, Chris J. and {Iyer}, Kartheik G. and {Noirot}, Ga{\"e}l and {Muzzin}, Adam and {Sawicki}, Marcin and {Brammer}, Gabriel and {Desprez}, Guillaume and {Rihtar{\v{s}}i{\v{c}}}, Gregor and {Zabl}, Johannes and {Abraham}, Roberto and {Brada{\v{c}}}, Maru{\v{s}}a and {Doyon}, Ren{\'e} and {Antwi-Danso}, Jacqueline and {Berek}, Samantha and {Brown}, Westley and {Estrada-Carpenter}, Vince and {Favaro}, Jeremy and {Felicioni}, Giordano and {Forrest}, Ben and {Gaspar}, Gaia and {Gould}, Katriona M.~L. and {Gledhill}, Rachel and {Harshan}, Anishya and {Jahan}, Nusrath and {Jagga}, Naadiyah and {Jude{\v{z}}}, Jon and {Marchesini}, Danilo and {Markov}, Vladan and {Matharu}, Jasleen and {MacFarland}, Shannon and {Merchant}, Maya and {M{\'e}rida}, Rosa M. and {Mowla}, Lamiya and {Myers}, Katherine and {Omori}, Kiyoaki C. and {Pacifici}, Camilla and {Ravindranath}, Swara and {Robbins}, Luke and {Strait}, Victoria and {Sok}, Visal and {Tan}, Vivian Yun Yan and {Tripodi}, Roberta and {Wilson}, Gillian and {Withers}, Sunna},
        title = "{CANUCS/Technicolor Data Release 1: Imaging, Photometry, Slit Spectroscopy, and Stellar Population Parameters}",
      journal = {\apjs},
         year = 2026,
        month = jan,
       volume = {282},
       number = {1},
          eid = {3},
        pages = {3},
          doi = {10.3847/1538-4365/ae1611},
archivePrefix = {arXiv},
       eprint = {2506.21685},
 primaryClass = {astro-ph.GA},
       adsurl = {https://ui.adsabs.harvard.edu/abs/2026ApJS..282....3S}
}

@ARTICLE{2025arXiv250918302P,
       author = {{Prieto-Lyon}, Gonzalo and {Mason}, Charlotte A. and {Strait}, Victoria and {Brammer}, Gabriel and {Naidu}, Rohan P. and {Meyer}, Romain A. and {Oesch}, Pascal and {Tacchella}, Sandro and {Covelo-Paz}, Alba and {Giovinazzo}, Emma and {Xiao}, Mengyuan},
        title = "{Lyman-alpha emission at the end of reionization: line strengths and profiles from MMT and JWST observations at z\raisebox{-0.5ex}\textasciitilde5-6}",
      journal = {arXiv e-prints},
         year = 2025,
        month = sep,
          eid = {arXiv:2509.18302},
        pages = {arXiv:2509.18302},
          doi = {10.48550/arXiv.2509.18302},
archivePrefix = {arXiv},
        note = {submitted to A\&A},
       eprint = {2509.18302},
 primaryClass = {astro-ph.GA},
       adsurl = {https://ui.adsabs.harvard.edu/abs/2025arXiv250918302P}
}

@ARTICLE{2025MNRAS.536...27W,
       author = {{Witstok}, Joris and {Maiolino}, Roberto and {Smit}, Renske and {Jones}, Gareth C. and {Bunker}, Andrew J. and {Helton}, Jakob M. and {Johnson}, Benjamin D. and {Tacchella}, Sandro and {Saxena}, Aayush and {Arribas}, Santiago and {Bhatawdekar}, Rachana and {Boyett}, Kristan and {Cameron}, Alex J. and {Cargile}, Phillip A. and {Carniani}, Stefano and {Charlot}, St{\'e}phane and {Chevallard}, Jacopo and {Curti}, Mirko and {Curtis-Lake}, Emma and {D'Eugenio}, Francesco and {Eisenstein}, Daniel J. and {Hainline}, Kevin and {Hausen}, Ryan and {Kumari}, Nimisha and {Laseter}, Isaac and {Maseda}, Michael V. and {Rieke}, Marcia and {Robertson}, Brant and {Scholtz}, Jan and {Shivaei}, Irene and {Williams}, Christina C. and {Willmer}, Christopher N.~A. and {Willott}, Chris},
        title = "{JADES: primaeval Lyman {\ensuremath{\alpha}} emitting galaxies reveal early sites of reionization out to redshift z \raisebox{-0.5ex}\textasciitilde 9}",
      journal = {\mnras},
         year = 2025,
        month = jan,
       volume = {536},
       number = {1},
        pages = {27-50},
          doi = {10.1093/mnras/stae2535},
archivePrefix = {arXiv},
       eprint = {2404.05724},
 primaryClass = {astro-ph.GA},
       adsurl = {https://ui.adsabs.harvard.edu/abs/2025MNRAS.536...27W}
}

@ARTICLE{2025MNRAS.542.2998H,
       author = {{Harvey}, Thomas and {Conselice}, Christopher J. and {Adams}, Nathan J. and {Austin}, Duncan and {Li}, Qiong and {Rusakov}, Vadim and {Westcott}, Lewi and {Goolsby}, Caio M. and {Lovell}, Christopher C. and {Cochrane}, Rachel K. and {Vijayan}, Aswin P. and {Trussler}, James},
        title = "{Behind the spotlight: a systematic assessment of outshining using NIRCam medium bands in the JADES Origins Field}",
      journal = {\mnras},
         year = 2025,
        month = oct,
       volume = {542},
       number = {4},
        pages = {2998-3027},
          doi = {10.1093/mnras/staf1396},
archivePrefix = {arXiv},
       eprint = {2504.05244},
 primaryClass = {astro-ph.GA},
       adsurl = {https://ui.adsabs.harvard.edu/abs/2025MNRAS.542.2998H}
}

@ARTICLE{2023ApJ...948..126G,
       author = {{Gim{\'e}nez-Arteaga}, Clara and {Oesch}, Pascal A. and {Brammer}, Gabriel B. and {Valentino}, Francesco and {Mason}, Charlotte A. and {Weibel}, Andrea and {Barrufet}, Laia and {Fujimoto}, Seiji and {Heintz}, Kasper E. and {Nelson}, Erica J. and {Strait}, Victoria B. and {Suess}, Katherine A. and {Gibson}, Justus},
        title = "{Spatially Resolved Properties of Galaxies at 5 < z < 9 in the SMACS 0723 JWST ERO Field}",
      journal = {\apj},
         year = 2023,
        month = may,
       volume = {948},
       number = {2},
          eid = {126},
        pages = {126},
          doi = {10.3847/1538-4357/acc5ea},
archivePrefix = {arXiv},
       eprint = {2212.08670},
 primaryClass = {astro-ph.GA},
       adsurl = {https://ui.adsabs.harvard.edu/abs/2023ApJ...948..126G}
}

@ARTICLE{2018MNRAS.476.1532S,
       author = {{Sorba}, Robert and {Sawicki}, Marcin},
        title = "{Spatially unresolved SED fitting can underestimate galaxy masses: a solution to the missing mass problem}",
      journal = {\mnras},
         year = 2018,
        month = may,
       volume = {476},
       number = {2},
        pages = {1532-1547},
          doi = {10.1093/mnras/sty186},
archivePrefix = {arXiv},
       eprint = {1801.07368},
 primaryClass = {astro-ph.GA},
       adsurl = {https://ui.adsabs.harvard.edu/abs/2018MNRAS.476.1532S}
}

@ARTICLE{2015MNRAS.452..235S,
       author = {{Sorba}, R. and {Sawicki}, M.},
        title = "{Missing stellar mass in SED fitting: spatially unresolved photometry can underestimate galaxy masses}",
      journal = {\mnras},
         year = 2015,
        month = sep,
       volume = {452},
       number = {1},
        pages = {235-245},
          doi = {10.1093/mnras/stv1235},
archivePrefix = {arXiv},
       eprint = {1506.01653},
 primaryClass = {astro-ph.GA},
       adsurl = {https://ui.adsabs.harvard.edu/abs/2015MNRAS.452..235S}
}

@ARTICLE{2010ApJ...724.1524O,
       author = {{Ono}, Yoshiaki and {Ouchi}, Masami and {Shimasaku}, Kazuhiro and {Dunlop}, James and {Farrah}, Duncan and {McLure}, Ross and {Okamura}, Sadanori},
        title = "{Stellar Populations of Ly{\ensuremath{\alpha}} Emitters at z \raisebox{-0.5ex}\textasciitilde 6-7: Constraints on the Escape Fraction of Ionizing Photons from Galaxy Building Blocks}",
      journal = {\apj},
         year = 2010,
        month = dec,
       volume = {724},
       number = {2},
        pages = {1524-1535},
          doi = {10.1088/0004-637X/724/2/1524},
archivePrefix = {arXiv},
       eprint = {1004.0963},
 primaryClass = {astro-ph.CO},
       adsurl = {https://ui.adsabs.harvard.edu/abs/2010ApJ...724.1524O}
}

@ARTICLE{1967ApJ...147..868P,
       author = {{Partridge}, R.~B. and {Peebles}, P.~J.~E.},
        title = "{Are Young Galaxies Visible?}",
      journal = {\apj},
         year = 1967,
        month = mar,
       volume = {147},
        pages = {868},
          doi = {10.1086/149079},
       adsurl = {https://ui.adsabs.harvard.edu/abs/1967ApJ...147..868P}
}

@ARTICLE{2003A&A...397..527S,
       author = {{Schaerer}, D.},
        title = "{The transition from Population III to normal galaxies: Lyalpha and He II emission and the ionising properties of high redshift starburst galaxies}",
      journal = {\aap},
         year = 2003,
        month = jan,
       volume = {397},
        pages = {527-538},
          doi = {10.1051/0004-6361:20021525},
archivePrefix = {arXiv},
       eprint = {astro-ph/0210462},
 primaryClass = {astro-ph},
       adsurl = {https://ui.adsabs.harvard.edu/abs/2003A&A...397..527S}
}

@ARTICLE{1999ApJ...518..138H,
       author = {{Haiman}, Zolt{\'a}n and {Spaans}, Marco},
        title = "{Models for Dusty Ly{\ensuremath{\alpha}} Emitters at High Redshift}",
      journal = {\apj},
         year = 1999,
        month = jun,
       volume = {518},
       number = {1},
        pages = {138-144},
          doi = {10.1086/307276},
archivePrefix = {arXiv},
       eprint = {astro-ph/9809223},
 primaryClass = {astro-ph},
       adsurl = {https://ui.adsabs.harvard.edu/abs/1999ApJ...518..138H}
}

@ARTICLE{2010Natur.468...49R,
       author = {{Robertson}, Brant E. and {Ellis}, Richard S. and {Dunlop}, James S. and {McLure}, Ross J. and {Stark}, Daniel P.},
        title = "{Early star-forming galaxies and the reionization of the Universe}",
      journal = {\nat},
         year = 2010,
        month = nov,
       volume = {468},
       number = {7320},
        pages = {49-55},
          doi = {10.1038/nature09527},
archivePrefix = {arXiv},
       eprint = {1011.0727},
 primaryClass = {astro-ph.CO},
       adsurl = {https://ui.adsabs.harvard.edu/abs/2010Natur.468...49R}
}

@ARTICLE{2022ApJ...935..167A,
       author = {{Astropy Collaboration} and {Price-Whelan}, Adrian M. and {Lim}, Pey Lian and {Earl}, Nicholas and {Starkman}, Nathaniel and {Bradley}, Larry and {Shupe}, David L. and {Patil}, Aarya A. and {Corrales}, Lia and {Brasseur}, C.~E. and {N{\"o}the}, Maximilian and {Donath}, Axel and {Tollerud}, Erik and {Morris}, Brett M. and {Ginsburg}, Adam and {Vaher}, Eero and {Weaver}, Benjamin A. and {Tocknell}, James and {Jamieson}, William and {van Kerkwijk}, Marten H. and {Robitaille}, Thomas P. and {Merry}, Bruce and {Bachetti}, Matteo and {G{\"u}nther}, H. Moritz and {Aldcroft}, Thomas L. and {Alvarado-Montes}, Jaime A. and {Archibald}, Anne M. and {B{\'o}di}, Attila and {Bapat}, Shreyas and {Barentsen}, Geert and {Baz{\'a}n}, Juanjo and {Biswas}, Manish and {Boquien}, M{\'e}d{\'e}ric and {Burke}, D.~J. and {Cara}, Daria and {Cara}, Mihai and {Conroy}, Kyle E. and {Conseil}, Simon and {Craig}, Matthew W. and {Cross}, Robert M. and {Cruz}, Kelle L. and {D'Eugenio}, Francesco and {Dencheva}, Nadia and {Devillepoix}, Hadrien A.~R. and {Dietrich}, J{\"o}rg P. and {Eigenbrot}, Arthur Davis and {Erben}, Thomas and {Ferreira}, Leonardo and {Foreman-Mackey}, Daniel and {Fox}, Ryan and {Freij}, Nabil and {Garg}, Suyog and {Geda}, Robel and {Glattly}, Lauren and {Gondhalekar}, Yash and {Gordon}, Karl D. and {Grant}, David and {Greenfield}, Perry and {Groener}, Austen M. and {Guest}, Steve and {Gurovich}, Sebastian and {Handberg}, Rasmus and {Hart}, Akeem and {Hatfield-Dodds}, Zac and {Homeier}, Derek and {Hosseinzadeh}, Griffin and {Jenness}, Tim and {Jones}, Craig K. and {Joseph}, Prajwel and {Kalmbach}, J. Bryce and {Karamehmetoglu}, Emir and {Ka{\l}uszy{\'n}ski}, Miko{\l}aj and {Kelley}, Michael S.~P. and {Kern}, Nicholas and {Kerzendorf}, Wolfgang E. and {Koch}, Eric W. and {Kulumani}, Shankar and {Lee}, Antony and {Ly}, Chun and {Ma}, Zhiyuan and {MacBride}, Conor and {Maljaars}, Jakob M. and {Muna}, Demitri and {Murphy}, N.~A. and {Norman}, Henrik and {O'Steen}, Richard and {Oman}, Kyle A. and {Pacifici}, Camilla and {Pascual}, Sergio and {Pascual-Granado}, J. and {Patil}, Rohit R. and {Perren}, Gabriel I. and {Pickering}, Timothy E. and {Rastogi}, Tanuj and {Roulston}, Benjamin R. and {Ryan}, Daniel F. and {Rykoff}, Eli S. and {Sabater}, Jose and {Sakurikar}, Parikshit and {Salgado}, Jes{\'u}s and {Sanghi}, Aniket and {Saunders}, Nicholas and {Savchenko}, Volodymyr and {Schwardt}, Ludwig and {Seifert-Eckert}, Michael and {Shih}, Albert Y. and {Jain}, Anany Shrey and {Shukla}, Gyanendra and {Sick}, Jonathan and {Simpson}, Chris and {Singanamalla}, Sudheesh and {Singer}, Leo P. and {Singhal}, Jaladh and {Sinha}, Manodeep and {Sip{\H{o}}cz}, Brigitta M. and {Spitler}, Lee R. and {Stansby}, David and {Streicher}, Ole and {{\v{S}}umak}, Jani and {Swinbank}, John D. and {Taranu}, Dan S. and {Tewary}, Nikita and {Tremblay}, Grant R. and {de Val-Borro}, Miguel and {Van Kooten}, Samuel J. and {Vasovi{\'c}}, Zlatan and {Verma}, Shresth and {de Miranda Cardoso}, Jos{\'e} Vin{\'\i}cius and {Williams}, Peter K.~G. and {Wilson}, Tom J. and {Winkel}, Benjamin and {Wood-Vasey}, W.~M. and {Xue}, Rui and {Yoachim}, Peter and {Zhang}, Chen and {Zonca}, Andrea and {Astropy Project Contributors}},
        title = "{The Astropy Project: Sustaining and Growing a Community-oriented Open-source Project and the Latest Major Release (v5.0) of the Core Package}",
      journal = {\apj},
         year = 2022,
        month = aug,
       volume = {935},
       number = {2},
          eid = {167},
        pages = {167},
          doi = {10.3847/1538-4357/ac7c74},
archivePrefix = {arXiv},
       eprint = {2206.14220},
 primaryClass = {astro-ph.IM},
       adsurl = {https://ui.adsabs.harvard.edu/abs/2022ApJ...935..167A}
}

@ARTICLE{2023ApJ...951..100N,
       author = {{Narayanan}, Desika and {Smith}, J. -D.~T. and {Hensley}, Brandon S. and {Li}, Qi and {Hu}, Chia-Yu and {Sandstrom}, Karin and {Torrey}, Paul and {Vogelsberger}, Mark and {Marinacci}, Federico and {Sales}, Laura V.},
        title = "{A Framework for Modeling Polycyclic Aromatic Hydrocarbon Emission in Galaxy Evolution Simulations}",
      journal = {\apj},
         year = 2023,
        month = jul,
       volume = {951},
       number = {2},
          eid = {100},
        pages = {100},
          doi = {10.3847/1538-4357/accf8d},
archivePrefix = {arXiv},
       eprint = {2301.07136},
 primaryClass = {astro-ph.GA},
       adsurl = {https://ui.adsabs.harvard.edu/abs/2023ApJ...951..100N}
}

@ARTICLE{2025MNRAS.544.4390N,
       author = {{Nakazato}, Yurina and {Ferrara}, Andrea},
        title = "{Radiation-driven dusty outflows from early galaxies}",
      journal = {\mnras},
         year = 2025,
        month = dec,
       volume = {544},
       number = {4},
        pages = {4390-4402},
          doi = {10.1093/mnras/staf2028},
archivePrefix = {arXiv},
       eprint = {2412.07598},
 primaryClass = {astro-ph.GA},
       adsurl = {https://ui.adsabs.harvard.edu/abs/2025MNRAS.544.4390N}
}

@ARTICLE{2023A&A...678A.174G,
       author = {{Goovaerts}, I. and {Pello}, R. and {Thai}, T.~T. and {Tuan-Anh}, P. and {Richard}, J. and {Claeyssens}, A. and {Carinos}, E. and {de la Vieuville}, G. and {Matthee}, J.},
        title = "{Evolution of the Lyman-{\ensuremath{\alpha}}-emitting fraction and UV properties of lensed star-forming galaxies in the range 2.9 < z < 6.7}",
      journal = {\aap},
         year = 2023,
        month = oct,
       volume = {678},
          eid = {A174},
        pages = {A174},
          doi = {10.1051/0004-6361/202347110},
archivePrefix = {arXiv},
       eprint = {2307.15559},
 primaryClass = {astro-ph.GA},
       adsurl = {https://ui.adsabs.harvard.edu/abs/2023A&A...678A.174G}
}

@misc{ormerod2025detection2175aauvbump,
      title={Detection of the 2175{\AA} UV Bump at z>7: Evidence for Rapid Dust Evolution in a Merging Reionisation-Era Galaxy}, 
      author={Katherine Ormerod and Joris Witstok and Renske Smit and Anna de Graaff and Jakob M. Helton and Michael V. Maseda and Irene Shivaei and Andrew J. Bunker and Stefano Carniani and Francesco D'Eugenio and Rachana Bhatawdekar and Jacopo Chevallard and Marijn Franx and Nimisha Kumari and Roberto Maiolino and Pierluigi Rinaldi and Brant Robertson and Sandro Tacchella},
      year={2025},
      eprint={2502.21119},
      archivePrefix={arXiv},
      primaryClass={astro-ph.GA},
      url={https://arxiv.org/abs/2502.21119}, 
}

@ARTICLE{2019ApJ...876....3L,
       author = {{Leja}, Joel and {Carnall}, Adam C. and {Johnson}, Benjamin D. and {Conroy}, Charlie and {Speagle}, Joshua S.},
        title = "{How to Measure Galaxy Star Formation Histories. II. Nonparametric Models}",
      journal = {\apj},
         year = 2019,
        month = may,
       volume = {876},
       number = {1},
          eid = {3},
        pages = {3},
          doi = {10.3847/1538-4357/ab133c},
archivePrefix = {arXiv},
       eprint = {1811.03637},
 primaryClass = {astro-ph.GA},
       adsurl = {https://ui.adsabs.harvard.edu/abs/2019ApJ...876....3L}
}

@ARTICLE{2016MNRAS.462.1415C,
       author = {{Chevallard}, Jacopo and {Charlot}, St{\'e}phane},
        title = "{Modelling and interpreting spectral energy distributions of galaxies with BEAGLE}",
      journal = {\mnras},
         year = 2016,
        month = oct,
       volume = {462},
       number = {2},
        pages = {1415-1443},
          doi = {10.1093/mnras/stw1756},
archivePrefix = {arXiv},
       eprint = {1603.03037},
 primaryClass = {astro-ph.GA},
       adsurl = {https://ui.adsabs.harvard.edu/abs/2016MNRAS.462.1415C}
}

@ARTICLE{2018MNRAS.480.4379C,
       author = {{Carnall}, A.~C. and {McLure}, R.~J. and {Dunlop}, J.~S. and {Dav{\'e}}, R.},
        title = "{Inferring the star formation histories of massive quiescent galaxies with BAGPIPES: evidence for multiple quenching mechanisms}",
      journal = {\mnras},
         year = 2018,
        month = nov,
       volume = {480},
       number = {4},
        pages = {4379-4401},
          doi = {10.1093/mnras/sty2169},
archivePrefix = {arXiv},
       eprint = {1712.04452},
 primaryClass = {astro-ph.GA},
       adsurl = {https://ui.adsabs.harvard.edu/abs/2018MNRAS.480.4379C}
}

@ARTICLE{2015ApJ...806..259R,
       author = {{Reddy}, Naveen A. and {Kriek}, Mariska and {Shapley}, Alice E. and {Freeman}, William R. and {Siana}, Brian and {Coil}, Alison L. and {Mobasher}, Bahram and {Price}, Sedona H. and {Sanders}, Ryan L. and {Shivaei}, Irene},
        title = "{The MOSDEF Survey: Measurements of Balmer Decrements and the Dust Attenuation Curve at Redshifts z \raisebox{-0.5ex}\textasciitilde 1.4-2.6}",
      journal = {\apj},
         year = 2015,
        month = jun,
       volume = {806},
       number = {2},
          eid = {259},
        pages = {259},
          doi = {10.1088/0004-637X/806/2/259},
archivePrefix = {arXiv},
       eprint = {1504.02782},
 primaryClass = {astro-ph.GA},
       adsurl = {https://ui.adsabs.harvard.edu/abs/2015ApJ...806..259R}
}

@ARTICLE{2016ApJS..227....2S,
       author = {{Salim}, Samir and {Lee}, Janice C. and {Janowiecki}, Steven and {da Cunha}, Elisabete and {Dickinson}, Mark and {Boquien}, M{\'e}d{\'e}ric and {Burgarella}, Denis and {Salzer}, John J. and {Charlot}, St{\'e}phane},
        title = "{GALEX-SDSS-WISE Legacy Catalog (GSWLC): Star Formation Rates, Stellar Masses, and Dust Attenuations of 700,000 Low-redshift Galaxies}",
      journal = {\apjs},
         year = 2016,
        month = nov,
       volume = {227},
       number = {1},
          eid = {2},
        pages = {2},
          doi = {10.3847/0067-0049/227/1/2},
archivePrefix = {arXiv},
       eprint = {1610.00712},
 primaryClass = {astro-ph.GA},
       adsurl = {https://ui.adsabs.harvard.edu/abs/2016ApJS..227....2S}
}

@ARTICLE{2020MNRAS.495.3602W,
       author = {{Whitler}, Lily R. and {Mason}, Charlotte A. and {Ren}, Keven and {Dijkstra}, Mark and {Mesinger}, Andrei and {Pentericci}, Laura and {Trenti}, Michele and {Treu}, Tommaso},
        title = "{The impact of scatter in the galaxy UV luminosity to halo mass relation on Ly {\ensuremath{\alpha}} visibility during the epoch of reionization}",
      journal = {\mnras},
         year = 2020,
        month = jul,
       volume = {495},
       number = {4},
        pages = {3602-3613},
          doi = {10.1093/mnras/staa1178},
archivePrefix = {arXiv},
       eprint = {1911.03499},
 primaryClass = {astro-ph.CO},
       adsurl = {https://ui.adsabs.harvard.edu/abs/2020MNRAS.495.3602W}
}

@ARTICLE{1965ApJ...142.1633G,
       author = {{Gunn}, James E. and {Peterson}, Bruce A.},
        title = "{On the Density of Neutral Hydrogen in Intergalactic Space.}",
      journal = {\apj},
         year = 1965,
        month = nov,
       volume = {142},
        pages = {1633-1636},
          doi = {10.1086/148444},
       adsurl = {https://ui.adsabs.harvard.edu/abs/1965ApJ...142.1633G}
}

@software{2022zndo...7299500B,
       author = {{Brammer}, Gabriel},
        title = "{msaexp: NIRSpec analyis tools}",
         year = 2023,
        month = sep,
          eid = {10.5281/zenodo.7299500},
          doi = {10.5281/zenodo.7299500},
      version = {0.6.17},
    publisher = {Zenodo},
       adsurl = {https://ui.adsabs.harvard.edu/abs/2022zndo...7299500B}
}

@ARTICLE{2024Sci...384..890H,
       author = {{Heintz}, Kasper E. and {Watson}, Darach and {Brammer}, Gabriel and {Vejlgaard}, Simone and {Hutter}, Anne and {Strait}, Victoria B. and {Matthee}, Jorryt and {Oesch}, Pascal A. and {Jakobsson}, P{\'a}ll and {Tanvir}, Nial R. and {Laursen}, Peter and {Naidu}, Rohan P. and {Mason}, Charlotte A. and {Killi}, Meghana and {Jung}, Intae and {Hsiao}, Tiger Yu-Yang and {Abdurro'uf} and {Coe}, Dan and {Arrabal Haro}, Pablo and {Finkelstein}, Steven L. and {Toft}, Sune},
        title = "{Strong damped Lyman-{\ensuremath{\alpha}} absorption in young star-forming galaxies at redshifts 9 to 11}",
      journal = {Science},
         year = 2024,
        month = may,
       volume = {384},
       number = {6698},
        pages = {890-894},
          doi = {10.1126/science.adj0343},
archivePrefix = {arXiv},
       eprint = {2306.00647},
 primaryClass = {astro-ph.GA},
       adsurl = {https://ui.adsabs.harvard.edu/abs/2024Sci...384..890H}
}

@ARTICLE{2025A&A...697A.189D,
       author = {{de Graaff}, Anna and {Brammer}, Gabriel and {Weibel}, Andrea and {Lewis}, Zach and {Maseda}, Michael V. and {Oesch}, Pascal A. and {Bezanson}, Rachel and {Boogaard}, Leindert A. and {Cleri}, Nikko J. and {Cooper}, Olivia R. and {Gottumukkala}, Rashmi and {Greene}, Jenny E. and {Hirschmann}, Michaela and {Hviding}, Raphael E. and {Katz}, Harley and {Labb{\'e}}, Ivo and {Leja}, Joel and {Matthee}, Jorryt and {McConachie}, Ian and {Miller}, Tim B. and {Naidu}, Rohan P. and {Price}, Sedona H. and {Rix}, Hans-Walter and {Setton}, David J. and {Suess}, Katherine A. and {Wang}, Bingjie and {Whitaker}, Katherine E. and {Williams}, Christina C.},
        title = "{RUBIES: A complete census of the bright and red distant Universe with JWST/NIRSpec}",
      journal = {\aap},
         year = 2025,
        month = may,
       volume = {697},
          eid = {A189},
        pages = {A189},
          doi = {10.1051/0004-6361/202452186},
archivePrefix = {arXiv},
       eprint = {2409.05948},
 primaryClass = {astro-ph.GA},
       adsurl = {https://ui.adsabs.harvard.edu/abs/2025A&A...697A.189D}
}

@ARTICLE{2025A&A...702A..33M,
       author = {{Markov}, V. and {Gallerani}, S. and {Pallottini}, A. and {Brada{\v{c}}}, M. and {Carniani}, S. and {Tripodi}, R. and {Noirot}, G. and {Di Mascia}, F. and {Parlanti}, E. and {Martis}, N.},
        title = "{Unveiling the trends between dust attenuation and galaxy properties at z {\ensuremath{\sim}} 2‑12 with the James Webb Space Telescope}",
      journal = {\aap},
         year = 2025,
        month = oct,
       volume = {702},
          eid = {A33},
        pages = {A33},
          doi = {10.1051/0004-6361/202555182},
archivePrefix = {arXiv},
       eprint = {2504.12378},
 primaryClass = {astro-ph.GA},
       adsurl = {https://ui.adsabs.harvard.edu/abs/2025A&A...702A..33M}
}

@ARTICLE{2024MNRAS.532..577E,
       author = {{Estrada-Carpenter}, Vicente and {Sawicki}, Marcin and {Brammer}, Gabe and {Desprez}, Guillaume and {Abraham}, Roberto and {Asada}, Yoshihisa and {Brada{\v{c}}}, Maru{\v{s}}a and {Iyer}, Kartheik G. and {Martis}, Nicholas S. and {Matharu}, Jasleen and {Mowla}, Lamiya and {Muzzin}, Adam and {Noirot}, Ga{\"e}l and {Sarrouh}, Ghassan T.~E. and {Strait}, Victoria and {Willott}, Chris J.},
        title = "{When, where, and how star formation happens in a galaxy pair at cosmic noon using CANUCS JWST/NIRISS grism spectroscopy}",
      journal = {\mnras},
         year = 2024,
        month = jul,
       volume = {532},
       number = {1},
        pages = {577-591},
          doi = {10.1093/mnras/stae1368},
archivePrefix = {arXiv},
       eprint = {2406.15551},
 primaryClass = {astro-ph.GA},
       adsurl = {https://ui.adsabs.harvard.edu/abs/2024MNRAS.532..577E}
}

@ARTICLE{2025ApJ...991..188E,
       author = {{Estrada-Carpenter}, Vicente and {Sawicki}, Marcin and {Abraham}, Roberto and {Asada}, Yoshihisa and {Brada{\v{c}}}, Maru{\v{s}}a and {Brammer}, Gabe and {Desprez}, Guillaume and {Iyer}, Kartheik G. and {Martis}, Nicholas S. and {Muzzin}, Adam and {Noirot}, Ga{\"e}l and {Rihtar{\v{s}}i{\v{c}}}, Gregor and {Sarrouh}, Ghassan T.~E. and {Willott}, Chris J. and {Favaro}, Jeremy and {Markov}, Vladan and {M{\'e}rida}, Rosa M. and {Myers}, Katherine and {Sok}, Visal},
        title = "{Metal-poor Star-forming Clumps in Cosmic Noon Galaxies: Evidence for Gas Inflow and Chemical Dilution Using JWST NIRISS}",
      journal = {\apj},
         year = 2025,
        month = oct,
       volume = {991},
       number = {2},
          eid = {188},
        pages = {188},
          doi = {10.3847/1538-4357/adfb64},
archivePrefix = {arXiv},
       eprint = {2508.00985},
 primaryClass = {astro-ph.GA},
       adsurl = {https://ui.adsabs.harvard.edu/abs/2025ApJ...991..188E}
}

@ARTICLE{2023A&A...679A..12M,
       author = {{Markov}, V. and {Gallerani}, S. and {Pallottini}, A. and {Sommovigo}, L. and {Carniani}, S. and {Ferrara}, A. and {Parlanti}, E. and {Di Mascia}, F.},
        title = "{Dust attenuation law in JWST galaxies at z {\ensuremath{\sim}} 7-8}",
      journal = {\aap},
         year = 2023,
        month = nov,
       volume = {679},
          eid = {A12},
        pages = {A12},
          doi = {10.1051/0004-6361/202346723},
       adsurl = {https://ui.adsabs.harvard.edu/abs/2023A&A...679A..12M}
}

@ARTICLE{Kneib96,
       author = {{Kneib}, J. -P. and {Ellis}, R.~S. and {Smail}, I. and {Couch}, W.~J. and {Sharples}, R.~M.},
        title = "{Hubble Space Telescope Observations of the Lensing Cluster Abell 2218}",
      journal = {\apj},
         year = 1996,
        month = nov,
       volume = {471},
        pages = {643},
          doi = {10.1086/177995},
archivePrefix = {arXiv},
       eprint = {astro-ph/9511015},
 primaryClass = {astro-ph},
       adsurl = {https://ui.adsabs.harvard.edu/abs/1996ApJ...471..643K}
}

@ARTICLE{2013ApJ...775L..16K,
       author = {{Kriek}, Mariska and {Conroy}, Charlie},
        title = "{The Dust Attenuation Law in Distant Galaxies: Evidence for Variation with Spectral Type}",
      journal = {\apjl},
         year = 2013,
        month = sep,
       volume = {775},
       number = {1},
          eid = {L16},
        pages = {L16},
          doi = {10.1088/2041-8205/775/1/L16},
archivePrefix = {arXiv},
       eprint = {1308.1099},
 primaryClass = {astro-ph.CO},
       adsurl = {https://ui.adsabs.harvard.edu/abs/2013ApJ...775L..16K}
}

@ARTICLE{2025NatCo..16.9830T,
       author = {{Tripodi}, Roberta and {Martis}, Nicholas and {Markov}, Vladan and {Brada{\v{c}}}, Maru{\v{s}}a and {Di Mascia}, Fabio and {Cammelli}, Vieri and {D'Eugenio}, Francesco and {Willott}, Chris and {Curti}, Mirko and {Bhatt}, Maulik and {Gallerani}, Simona and {Rihtar{\v{s}}i{\v{c}}}, Gregor and {Singh}, Jasbir and {Gaspar}, Gaia and {Harshan}, Anishya and {Jude{\v{z}}}, Jon and {Merida}, Rosa M. and {Desprez}, Guillaume and {Sawicki}, Marcin and {Goovaerts}, Ilias and {Muzzin}, Adam and {Noirot}, Ga{\"e}l and {Sarrouh}, Ghassan T.~E. and {Abraham}, Roberto and {Asada}, Yoshihisa and {Brammer}, Gabriel and {Estrada-Carpenter}, Vicente and {Felicioni}, Giordano and {Fujimoto}, Seiji and {Iyer}, Kartheik and {Mowla}, Lamiya and {Strait}, Victoria},
        title = "{Extreme properties of a compact and massive accreting black hole host in the first 500 Myr}",
      journal = {Nature Communications},
         year = 2025,
        month = nov,
       volume = {16},
       number = {1},
          eid = {9830},
        pages = {9830},
          doi = {10.1038/s41467-025-65070-x},
archivePrefix = {arXiv},
       eprint = {2412.04983},
 primaryClass = {astro-ph.GA},
       adsurl = {https://ui.adsabs.harvard.edu/abs/2025NatCo..16.9830T}
}

@ARTICLE{2020ApJ...888..108B,
       author = {{Battisti}, A.~J. and {Cunha}, E. da and {Shivaei}, I. and {Calzetti}, D.},
        title = "{The Strength of the 2175 {\r{A}} Feature in the Attenuation Curves of Galaxies at 0.1 < z {\ensuremath{\lesssim}} 3}",
      journal = {\apj},
         year = 2020,
        month = jan,
       volume = {888},
       number = {2},
          eid = {108},
        pages = {108},
          doi = {10.3847/1538-4357/ab5fdd},
archivePrefix = {arXiv},
       eprint = {1912.05206},
 primaryClass = {astro-ph.GA},
       adsurl = {https://ui.adsabs.harvard.edu/abs/2020ApJ...888..108B}
}

@BOOK{2006agna.book.....O,
       author = {{Osterbrock}, Donald E. and {Ferland}, Gary J.},
        title = "{Astrophysics of gaseous nebulae and active galactic nuclei}",
         year = 2006,
       adsurl = {https://ui.adsabs.harvard.edu/abs/2006agna.book.....O}
}

@ARTICLE{2008ApJ...685.1046L,
       author = {{Li}, Aigen and {Liang}, S.~L. and {Kann}, D.~A. and {Wei}, D.~M. and {Klose}, S. and {Wang}, Y.~J.},
        title = "{On Dust Extinction of Gamma-Ray Burst Host Galaxies}",
      journal = {\apj},
         year = 2008,
        month = oct,
       volume = {685},
       number = {2},
        pages = {1046-1051},
          doi = {10.1086/591049},
archivePrefix = {arXiv},
       eprint = {0808.4115},
 primaryClass = {astro-ph},
       adsurl = {https://ui.adsabs.harvard.edu/abs/2008ApJ...685.1046L}
}

@ARTICLE{2017RMxAA..53..385F,
       author = {{Ferland}, G.~J. and {Chatzikos}, M. and {Guzm{\'a}n}, F. and {Lykins}, M.~L. and {van Hoof}, P.~A.~M. and {Williams}, R.~J.~R. and {Abel}, N.~P. and {Badnell}, N.~R. and {Keenan}, F.~P. and {Porter}, R.~L. and {Stancil}, P.~C.},
        title = "{The 2017 Release Cloudy}",
      journal = {\rmxaa},
         year = 2017,
        month = oct,
       volume = {53},
        pages = {385-438},
          doi = {10.48550/arXiv.1705.10877},
archivePrefix = {arXiv},
       eprint = {1705.10877},
 primaryClass = {astro-ph.GA},
       adsurl = {https://ui.adsabs.harvard.edu/abs/2017RMxAA..53..385F}
}

@ARTICLE{2025arXiv251018946N,
       author = {{Neyer}, Meredith and {Smith}, Aaron and {Vogelsberger}, Mark and {{\'A}ngela Garc{\'\i}a}, Luz and {Kannan}, Rahul and {Garaldi}, Enrico and {Keating}, Laura},
        title = "{The THESAN project: Lyman-alpha emitters as probes of ionized bubble sizes}",
      journal = {arXiv e-prints},
         year = 2025,
        month = oct,
          eid = {arXiv:2510.18946},
        pages = {arXiv:2510.18946},
          doi = {10.48550/arXiv.2510.18946},
archivePrefix = {arXiv},
       eprint = {2510.18946},
 primaryClass = {astro-ph.GA},
       adsurl = {https://ui.adsabs.harvard.edu/abs/2025arXiv251018946N}
}

@ARTICLE{2019ApJ...879..116I,
       author = {{Iyer}, Kartheik G. and {Gawiser}, Eric and {Faber}, Sandra M. and {Ferguson}, Henry C. and {Kartaltepe}, Jeyhan and {Koekemoer}, Anton M. and {Pacifici}, Camilla and {Somerville}, Rachel S.},
        title = "{Nonparametric Star Formation History Reconstruction with Gaussian Processes. I. Counting Major Episodes of Star Formation}",
      journal = {\apj},
         year = 2019,
        month = jul,
       volume = {879},
       number = {2},
          eid = {116},
        pages = {116},
          doi = {10.3847/1538-4357/ab2052},
archivePrefix = {arXiv},
       eprint = {1901.02877},
 primaryClass = {astro-ph.GA},
       adsurl = {https://ui.adsabs.harvard.edu/abs/2019ApJ...879..116I}
}

@ARTICLE{2003ApJ...588...65S,
       author = {{Shapley}, Alice E. and {Steidel}, Charles C. and {Pettini}, Max and {Adelberger}, Kurt L.},
        title = "{Rest-Frame Ultraviolet Spectra of z\raisebox{-0.5ex}\textasciitilde3 Lyman Break Galaxies}",
      journal = {\apj},
         year = 2003,
        month = may,
       volume = {588},
       number = {1},
        pages = {65-89},
          doi = {10.1086/373922},
archivePrefix = {arXiv},
       eprint = {astro-ph/0301230},
 primaryClass = {astro-ph},
       adsurl = {https://ui.adsabs.harvard.edu/abs/2003ApJ...588...65S}
}

@ARTICLE{2014PASA...31...40D,
       author = {{Dijkstra}, Mark},
        title = "{Ly{\ensuremath{\alpha}} Emitting Galaxies as a Probe of Reionisation}",
      journal = {\pasa},
         year = 2014,
        month = oct,
       volume = {31},
          eid = {e040},
        pages = {e040},
          doi = {10.1017/pasa.2014.33},
archivePrefix = {arXiv},
       eprint = {1406.7292},
 primaryClass = {astro-ph.CO},
       adsurl = {https://ui.adsabs.harvard.edu/abs/2014PASA...31...40D}
}

@ARTICLE{1999ApJ...521...64M,
       author = {{Meurer}, Gerhardt R. and {Heckman}, Timothy M. and {Calzetti}, Daniela},
        title = "{Dust Absorption and the Ultraviolet Luminosity Density at z \raisebox{-0.5ex}\textasciitilde 3 as Calibrated by Local Starburst Galaxies}",
      journal = {\apj},
         year = 1999,
        month = aug,
       volume = {521},
       number = {1},
        pages = {64-80},
          doi = {10.1086/307523},
archivePrefix = {arXiv},
       eprint = {astro-ph/9903054},
 primaryClass = {astro-ph},
       adsurl = {https://ui.adsabs.harvard.edu/abs/1999ApJ...521...64M}
}

@ARTICLE{2024ApJ...962...24S,
       author = {{Sanders}, Ryan L. and {Shapley}, Alice E. and {Topping}, Michael W. and {Reddy}, Naveen A. and {Brammer}, Gabriel B.},
        title = "{Direct T $_{e}$-based Metallicities of z = 2{\textendash}9 Galaxies with JWST/NIRSpec: Empirical Metallicity Calibrations Applicable from Reionization to Cosmic Noon}",
      journal = {\apj},
         year = 2024,
        month = feb,
       volume = {962},
       number = {1},
          eid = {24},
        pages = {24},
          doi = {10.3847/1538-4357/ad15fc},
archivePrefix = {arXiv},
       eprint = {2303.08149},
 primaryClass = {astro-ph.GA},
       adsurl = {https://ui.adsabs.harvard.edu/abs/2024ApJ...962...24S}
}

@ARTICLE{martis24,
       author = {{Martis}, Nicholas S. and {Sarrouh}, Ghassan T.~E. and {Willott}, Chris J. and {Abraham}, Roberto and {Asada}, Yoshihisa and {Brada{\v{c}}}, Maru{\v{s}}a and {Brammer}, Gabriel B. and {Desprez}, Guillaume and {Harshan}, Anishya and {Muzzin}, Adam and {Noirot}, Ga{\"e}l and {Rihtar{\v{s}}i{\v{c}}}, Gregor and {Sawicki}, Marcin and {Strait}, Victoria},
        title = "{Modeling and Subtracting Diffuse Cluster Light in JWST Images: A Relation between the Spatial Distribution of Globular Clusters, Dwarf Galaxies, and Intracluster Light in the Lensing Cluster SMACS 0723}",
      journal = {\apj},
         year = 2024,
        month = nov,
       volume = {975},
       number = {1},
          eid = {76},
        pages = {76},
          doi = {10.3847/1538-4357/ad7735},
archivePrefix = {arXiv},
       eprint = {2401.01945},
 primaryClass = {astro-ph.GA},
       adsurl = {https://ui.adsabs.harvard.edu/abs/2024ApJ...975...76M}
}

\begin{appendix}  
\onecolumn

\section{Overplotted spectra of individual slit regions (S1–S3)} \label{three_spectra}

\begin{figure}[h]
    \centering
    \includegraphics[width=\hsize]{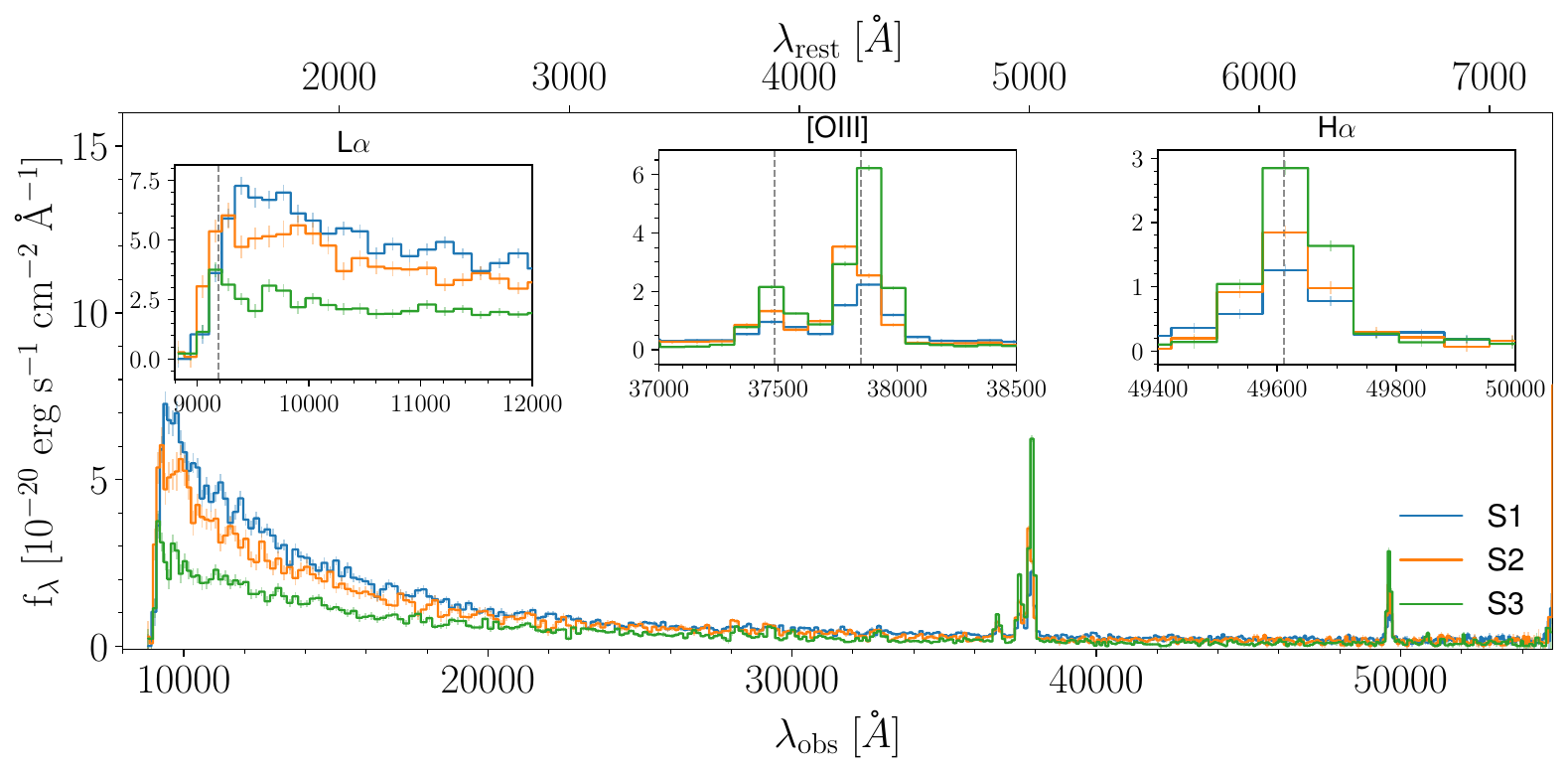}
\caption{{Photometry-rescaled NIRSpec PRISM slit spectra extracted from three adjacent slit regions (S1, S2, and S3; shown in blue, orange, and green, respectively). The main panel displays the spectra as a function of observed wavelength, with the corresponding rest-frame wavelength shown on the upper axis. Inset panels present zoom-ins around the Ly$\alpha$, [\ion{O}{III}] doublet, and H$\alpha$ emission-line regions (left, center, and right, respectively). Dashed vertical lines mark the expected wavelengths of the emission lines. Inset panels share the same spectral flux density $f_\lambda$ and observed wavelength $\lambda_{\rm obs}$ units as the main panel.
}
    \label{fig:three_spectra}}
\end{figure}

\FloatBarrier

\section{Integrated and slit-resolved Ly$\alpha$ fluxes}\label{Lya_flux}

In Table \ref{Lya_table} we report observed Ly$\alpha$ fluxes not corrected for gravitational lensing magnification, measured for the integrated system and within the three NIRSpec slit regions (S1–S3), using different methods.  Column (b) lists fluxes from the NIRISS Ly$\alpha$ map integrated within the adopted aperture encompassing the entire source (total) and the three slits (S1-S3), using \texttt{CARTA} (Cube Analysis and Rendering Tool for Astronomy; \citealp{2021zndo...3377984C}).
The sum of the map-based fluxes within S1–S3 corresponds to $\approx55\%$ of the total flux of the system, indicating that nearly half of the Ly$\alpha$ emission lies outside the NIRSpec apertures. This highlights that, for spatially extended Ly$\alpha$ emission, NIRSpec slit observations can recover only a fraction of the total flux, an effect that should be considered in large survey studies of LAEs.
Column (c) reports the integrated Ly$\alpha$ flux derived from a direct forward-model fit to the 2D NIRISS slitless spectrum using a template-based approach with \texttt{grizli}, where the flux corresponds to the best-fit scaling of the Ly$\alpha$ template.

Columns (d) and (e) list Ly$\alpha$ fluxes from fits to the photometry-rescaled 1D NIRSpec spectra (v4 and v3) using the Dawn JWST Archive (DJA) NIRSpec data products code \citep{2024Sci...384..890H, 2025A&A...697A.189D}.
A marginal Ly$\alpha$ detection ($\rm{S/N} \approx 4$) is present in the v3 NIRSpec spectrum of S3, while no significant line is recovered in the corresponding v4 extraction. This discrepancy originates from the reduction-dependent background treatment, potentially including self-subtraction effects in v4.

\begin{table}[h!]
\centering
\caption{Ly$\alpha$ fluxes $F_{\rm Ly\alpha}\ [\times10^{-17}\ \mathrm{erg\ cm^{-2}\ s^{-1}}]$ or their $3\sigma$ upper limits.}
\label{Lya_table}

\begin{tabular}{lccccc}
\hline\hline
\noalign{\smallskip}

Region (a) &
NIRISS Ly$\alpha$ map (b) &
NIRISS 2D spectrum (c) &
v4 NIRSpec 1D (d) &
v3 NIRSpec 1D (e) \\

\noalign{\smallskip}
\hline
\noalign{\smallskip}

Total & $7.0 \pm 0.3$ & $7.5 \pm 0.2$ & ... & ... \\
 \noalign{\smallskip}
S1    & $1.1 \pm 0.1$ & ... & $< 0.6$ & $< 0.9$ \\
 \noalign{\smallskip}
S2    & $1.4 \pm 0.1$ & ... & $< 0.8$ & $< 0.8$ \\
 \noalign{\smallskip}
S3    & $1.3 \pm 0.1$ & ... & $< 0.45$& $0.60 \pm 0.15$ \\

\noalign{\smallskip}
\hline
\end{tabular}
\tablefoot{The fluxes were measured for the integrated system and for the three NIRSpec slit regions (S1–S3) from the NIRISS Ly$\alpha$ map (b), the integrated NIRISS spectrum (c), and the v4 and v3 NIRSpec slit spectra (d and e, respectively). All Ly$\alpha$ fluxes are not corrected for gravitational lensing magnification.}
\end{table}

\end{appendix}
\end{document}